\documentclass[%
reprint,
twocolumn,
superscriptaddress,
preprintnumbers,
bibnotes,
amsmath,amssymb,
prl,
lengthcheck
]{revtex4-1}

\usepackage{graphicx}% Include figure files
\usepackage{dcolumn}% Align table columns on decimal point
\usepackage{bm}% bold math
\usepackage{dsfont}% bold math
\usepackage{newtxtext,newtxmath}
\usepackage[colorlinks,urlcolor=blue,linkcolor=blue,citecolor=blue,anchorcolor=blue]{hyperref}
\usepackage{color}
\begin{document}

\title{Exact Solutions Disentangle Higher-Order Topology in 2D Non-Hermitian Lattices}

\author{Lingfang Li}
\affiliation{Key Laboratory of Quantum Materials and Devices of Ministry of Education, School of Physics, Frontiers Science Center for Mobile Information Communication and Security, Southeast University, Nanjing 211189, China}

\author{Yating Wei}
\affiliation{Key Laboratory of Quantum Materials and Devices of Ministry of Education, School of Physics, Frontiers Science Center for Mobile Information Communication and Security, Southeast University, Nanjing 211189, China}

\author{Gangzhou Wu}
\affiliation{Key Laboratory of Quantum Materials and Devices of Ministry of Education, School of Physics, Frontiers Science Center for Mobile Information Communication and Security, Southeast University, Nanjing 211189, China}

\author{Yang Ruan}
\affiliation{Key Laboratory of Quantum Materials and Devices of Ministry of Education, School of Physics, Frontiers Science Center for Mobile Information Communication and Security, Southeast University, Nanjing 211189, China}

\author{Shihua Chen}
\email{cshua@seu.edu.cn}
\affiliation{Key Laboratory of Quantum Materials and Devices of Ministry of Education, School of Physics, Frontiers Science Center for Mobile Information Communication and Security, Southeast University, Nanjing 211189, China}
\affiliation{Purple Mountain Laboratories, Nanjing 211111, China}

\author{Ching Hua Lee}
\email{phylch@nus.edu.sg}
\affiliation{Department of Physics, National University of Singapore, Singapore 117551, Republic of Singapore}

\author{Zhenhua Ni}
\email{zhni@seu.edu.cn}
\affiliation{Key Laboratory of Quantum Materials and Devices of Ministry of Education, School of Physics, Frontiers Science Center for Mobile Information Communication and Security, Southeast University, Nanjing 211189, China}
\affiliation{Purple Mountain Laboratories, Nanjing 211111, China}

\date{\today}% It is always \today, today,

\begin{abstract}
We report the exact closed-form solutions for higher-order topological states as well as explicit energy-spectrum relationships in two-dimensional (2D) non-Hermitian multi-orbital lattices with generalized boundary conditions. These analytical solutions unequivocally confirm that topological edge states in a 2D non-Hermitian system which feature point-gap topology must undergo the non-Hermitian skin effect along the edge. Under double open boundary conditions, the occurrence of the non-Hermitian skin effect for either topological edge states or bulk states can be accurately predicted by our proposed winding numbers. We unveil that the zero-energy topological corner state only manifests itself on a corner where two nearby gapped edge states intersect, and thus can either disappear completely or strengthen drastically due to the non-Hermitian skin effect of gapped topological edge states. Our analytical results offer direct insight into the non-Bloch band topology in two or higher dimensions and trigger experimental investigations into many related phenomena such as quadrupole topological insulators and topological lasing.
\end{abstract}

\maketitle

%\tableofcontents

Topological band theory characterizes edge or corner states by means of the nontrivial topology and symmetry of the bulk band structure \cite{Bansil16,Scheurer20,Xie21,Trifun21,Slager13,Chiu16,Kruthoff17}. It has not only revolutionized our understanding of topological phases of matter such as topological insulators \cite{Roth09,Hsieh08,Qi11,Wang09}, but also opened an entirely new avenue of research into exotic functionalities such as topological microcomb generation \cite{Mittal21}, flat-band fractional topological phases \cite{Regnault11,Wu2012,Lee2016,Lee15,Behrmann16,Lee14}, topological sensing\cite{Zhang24,Budich20}, and robust quantum entanglement \cite{Alexandradinata84,Xue24,Gu16,Hughes11,Dai22}. This great success has its roots in that the involved topology can be well understood using an abundance of solvable tight-binding models, e.g., the Haldane model \cite{Haldane88,Li2022}, the Su-Schrieffer-Heeger (SSH) model \cite{Su79,Tak18,Par18,Ste21}, and the Hatano-Nelson (HN) model \cite{Hatano96,Gong18}, which can be implemented in a broad range of practical settings across different disciplines \cite{Shen13,Roushan14,Zhang19,Xue22,Imhof18,Lu14,Ozawa2019}, with various geometric structures \cite{Chen19,Zhou20,Kim20b,Lan2024,Lang12,ChenG20,Stegmaier24,Shang24,Wang23}.

However, for a wide diversity of two-dimensional (2D) systems \cite{Zhang05,Khan17,Xu2023,Xia23,Edvard22,ChenY22} or beyond \cite{Kim2020}, especially those featuring non-Hermiticity \cite{Ash20,Kawa19,Cou21,RYang22,Okuma23,Kun18,Long19,Ber21}, the insight into the underlying topology is quite limited, with many issues still open to debate. For example, in 1D non-Hermitian systems, it has turned out that the non-Hermitian skin effect (NHSE) \cite{Yao18,Alvarez18,LeeC19,Li2020,Borgnia20,Longhi19,Zhao2019,Wei20,Sun21,Bhargava21,Schindler21,MHL22,Fang22,Zhang22b,Lin23,
Zhou23,WWang23,Tai23,Arouca20,Rafi-Ul-Islam22,Gal23,Jana23,Yang22,Qin23,Shen24,Qin24,Li22} stems from the intrinsic point-gap topology \cite{Longhi22} defined by $W=(2\pi i)^{-1}\oint_0^{2\pi}dk \partial_k \mathrm{ln}(\mathrm{det}[H(k)-E_{\mathrm{OBC}}])$ \cite{Okuma2020,Zha20}, which can be visualized as the winding of the energy spectra of the Bloch Hamiltonian $H(k)$ around any base energy $E_{\mathrm{OBC}}$ obtained under open boundary condition (OBC). But in a fully open 2D non-Hermitian system, the above definition of topological invariant for 2D Bloch Hamiltonian $H(k_x,k_y)$ is no longer valid \cite{Kawabata20,Okugawa20,Jiang23,Lei24}. Then, how does one accurately determine the topological origin of the NHSE occurring in the bulk \cite{Shang22,Hou24} or on the edges \cite{Lee19,Zou21,Zhang21,Li20}? Where exactly does the topological in-gap corner state \cite{XieBY18,XieBY19,LiuZ24} arise and can it become stronger in 2D non-Hermitian systems? In recent years, there have been intensive theoretical studies on these issues, but they relied mainly on numerical calculations of relevant toy models \cite{LiuF17,LiuT19,LiuF19,Lee18}. While straightforward, such numerical solutions may often not lend sufficient insight into the intricate interplay between various physical ingredients involved.

In this work, we wish to address these issues on an analytical level, using two typical 2D non-Hermitian SSH models. Our exact closed-form solutions offer explicit relationships between the energy spectra under different boundary conditions, by which the topological invariants for NHSEs can be defined. We show that the topological edge states feature a point-gap topology as well \cite{Lee19} and that the zero-energy corner states only arise on the corner where the topological edge states on adjoining edges intersect, hence the name in-edge corner states. Because of NHSE, the topological edge states may redistribute on the edges, resulting in the complete obliteration or enhanced localization of in-edge corner states. This unusual property may find potential applications in design of quadrupole topological insulators \cite{Garcia18,Lv21} and edge topological lasing \cite{Harari18,Bandres18,Lu2024}

\begin{figure}[ht]
\begin{center}
	\includegraphics[width=8.6cm]{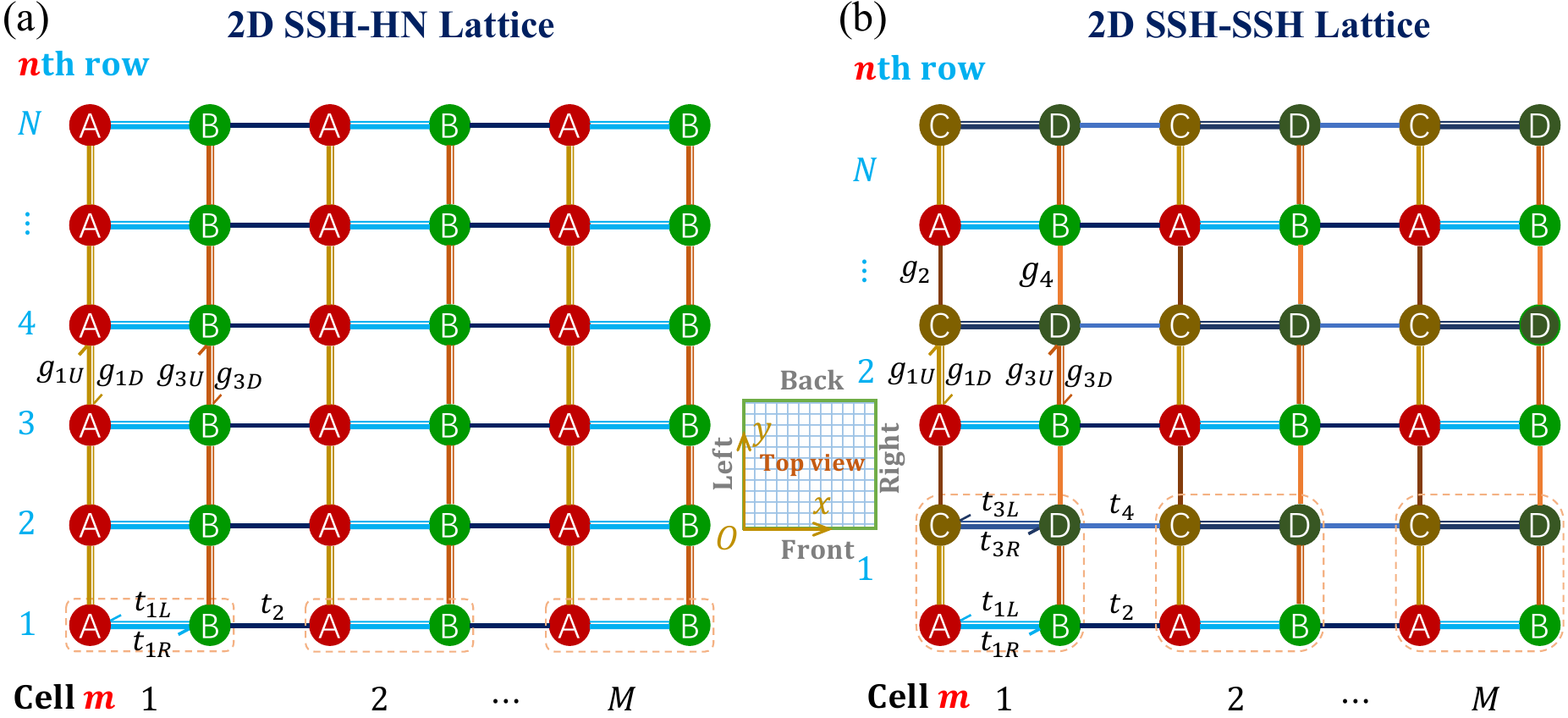}%%
	\caption{Schematic sketch of (a) a 2D non-Hermitian SSH-HN lattice and (b) a 2D non-Hermitian SSH-SSH lattice, where $t_{1L,R}$ and $t_{2}$ (similarly $t_{3L,R}$ and $t_{4}$) represent intracell and intercell hopping amplitudes in the horizontal chain, respectively, and $g_{1,3X}$ ($X=U,D$) and $g_{2,4}$ signify hoppings in the vertical chain. The inset in between shows the coordinate axes and lattice directions. } \label{fig1}
\end{center}
\end{figure}

Let us first consider a 2D lattice as sketched in Fig. \ref{fig1}(a), which can be deemed as a stack of 1D horizontal SSH lattice such that it is of HN type in the vertical direction \cite{Edvard22}. The Hamiltonian for such 2D SSH-HN lattice reads
\begin{eqnarray}
\hat{H}^{\mathrm{SN}}&=&\sum_{n,m}(\hat{C}^{\dagger}_{n,m}M_1\hat{C}_{n,m}+\hat{C}^{\dagger}_{n,m+1}M_{2}^{\dagger}\hat{C}_{n,m}
+\hat{C}^{\dagger}_{n,m}M_{2}\hat{C}_{n,m+1} \nonumber\\
&&+\hat{C}^{\dagger}_{n+1,m}M_{U}\hat{C}_{n,m}+\hat{C}^{\dagger}_{n,m}M_{D}\hat{C}_{n+1,m})+\hat{H}_{\mathrm{B1}}+\hat{H}_{\mathrm{B2}},\label{Eq1}
\end{eqnarray}
where $\hat{C}^{\dagger}_{n,m} = (\hat{a}^{\dagger}_{n,m}, \hat{b}^{\dagger}_{n,m})$ are the creation operators of particles on sublattices A and B of the cell in the $n$th row and $m$th column of lattice, and $M_{1,2}$ and $M_{X}$ ($X=U,D$) are the $2\times 2$ matrices given by
\begin{eqnarray}
M_{1}=	\left[\begin{array}{cc}
		0 &  t_{1L}  \\
		t_{1R} &0    \\		
	\end{array}\right],
M_{2}=	\left[\begin{array}{cc}
		0 &  0 \\
		t_{2} &  0  \\
	\end{array}\right],
M_{X}=	\left[\begin{array}{cc}
		g_{1X} &  0  \\
		0 &  g_{3X}   \\
	 \end{array}\right]. \label{Eq2}
\end{eqnarray}
Here, we use the parameters $t_{1L,R}$ ($t_{2}$) $\in\mathds{R}$ to represent the nonreciprocal \cite{Bran19} (reciprocal) hopping amplitudes  in the horizontal SSH chain, and use $g_{1X}$ and $g_{3X}$ for the hoppings between adjacent sites in the vertical HN chain.
The last two terms on the righthand side of Eq. (\ref{Eq1}), i.e.,
\begin{equation}
\hat{H}_{\mathrm{B1}}=\sum_{n}(\delta_1\hat{C}^{\dagger}_{n,1}M_{2}^{\dagger}
\hat{C}_{n,M}+\delta_2\hat{C}^{\dagger}_{n,M}M_{2}\hat{C}_{n,1}), \label{Eq3}
\end{equation}
\begin{equation}
\hat{H}_{\mathrm{B2}}=\sum_{m}(\kappa_1\hat{C}^{\dagger}_{1,m}M_{U}\hat{C}_{N,m}+\kappa_2\hat{C}^{\dagger}_{N,m}M_{D}\hat{C}_{1,m}), \label{Eq4}
\end{equation}
denote the boundary conditions imposed in the $x$ and $y$ dimensions, respectively.

A universal consideration of the boundary conditions entails the following three types: ({\it i}) double generalized boundary condition (dGBC) defined by $\delta_1=\delta_2^{-1}=e^{i\varphi}$ and $\kappa_1=\kappa_2^{-1}=e^{i\vartheta}$, with $\varphi,\vartheta\in \mathds{C}$, which includes the double periodic boundary condition (dPBC), where $\varphi=\vartheta=0$, as a special case \cite{Guo21}; ({\it ii}) unidirectional OBC, which can be subdivided into $x$OBC, where $\hat{H}_{\mathrm{B1}}=0$ and $\kappa_1=\kappa_2^{-1}=e^{i\vartheta}$, and $y$OBC, where $\delta_1=\delta_2^{-1}=e^{i\varphi}$ and $\hat{H}_{\mathrm{B2}}=0$; and ({\it iii}) double OBC (dOBC), which means $\hat{H}_{\mathrm{B1}}=\hat{H}_{\mathrm{B2}}=0$. Besides, we assume $g_{3X}=\mu g_{1X}$ ($\mu\in \mathds{R}$) so that Hamiltonian (\ref{Eq1}) has closed-form analytical solutions in all the above three boundary conditions (see Sec. I in Supplementary Material (SM) for details). Obviously, Hamiltonian (\ref{Eq1}) respects the time-reversal symmetry \cite{Kawa19,Li24}.  Therefore, the eigenenergies will come in $(E, -E^*)$ pairs for imaginary $g_{1X}$ or in $(E, E^*)$ pairs for real $g_{1X}$.

For the rectangular lattice of $2M$ sites in $x$ and $N$ sites in $y$ dimension, we can solve $\hat{H}^{\mathrm{SN}}|\psi\rangle=E|\psi\rangle$ via a gauge transform to get the energy spectrum under dGBC,
\begin{equation}
E^{\mathrm{dGBC}}=\frac{(\mu+1) P \pm\sqrt{(\mu-1)^2P^2+4T}}{2}, \label{Eq5}
\end{equation}
where
\begin{equation}
P=g_{1D} \omega_j+\frac{g_{1U}}{\omega_j},~~~T=(t_{1R}+t_{2}\varpi_\ell)(t_{1L}+\frac{t_{2}}{\varpi_\ell}), \label{Eq6}
\end{equation}
with $\varpi_\ell=\exp(\frac{i2\ell\pi}{M}-\frac{i\varphi}{M})$ and $\omega_j=\exp(\frac{i2j\pi}{N}-\frac{i\vartheta}{N})$. Here the subscripts $\ell=1,\cdots,M$ and $j=1,\cdots,N$ are used for labelling the eigenstates. The corresponding solutions for dPBC can follow easily by taking $\vartheta=\varphi=0$ therein.

On the other side, under $x$OBC, namely, the lattice entails OBC in the $x$ direction only, the energy spectrum would take
\begin{equation}
E^{x\mathrm{OBC}}=\frac{(\mu+1) P \pm\sqrt{(\mu-1)^2 P^2+4R}}{2}, \label{Eq7}
\end{equation}
where $P$ retains the same form as in Eq. (\ref{Eq5}), but $R$ replaces $T$, with the following form
\begin{equation}
R=2t_{2}\sqrt{t_{1L}t_{1R}}\cos(\theta_\ell)+t_{1L}t_{1R}+t_{2}^2. \label{Eq8}
\end{equation}
Here, $\theta_\ell$ is one of $M$ complex roots of the equation $\mathcal{T}(M+1)=0$, where
\begin{equation}
\mathcal{T}(m)=\frac{t_{2}\sin[(m-1)\theta_\ell]}{\sin(\theta_\ell)}+\frac{\sqrt{t_{1L}t_{1R}}\sin(m\theta_\ell)}{\sin(\theta_\ell)}. \label{Eq9}
\end{equation}
In a similar fashion, under $y$OBC, by replacing $P\rightarrow 2Q$ in Eq. (\ref{Eq5}) only, one can obtain the eigenenergy as
\begin{equation}
E^{y\mathrm{OBC}}=(\mu+1) Q \pm\sqrt{(\mu-1)^2 Q^2+T}, \label{Eq10}
\end{equation}
where
\begin{equation}
Q=\sqrt{g_{1D}g_{1U}}\cos{\phi_j},~~~\phi_j=\frac{j\pi}{N+1}. \label{Eq11}
\end{equation}

Remarkably, when making substitutions $P\rightarrow 2Q$ and $T\rightarrow R$ simultaneously in Eq. (\ref{Eq5}), one obtains
\begin{equation}
E^{\mathrm{dOBC}}=(\mu+1)Q\pm\sqrt{(\mu-1)^2 Q^2+R}, \label{Eq12}
\end{equation}
which is nothing but the energy spectrum of Hamiltonian (\ref{Eq1}) for dOBC, namely, $\hat{H}_{\mathrm{B1}}=\hat{H}_{\mathrm{B2}}=0$. Then, the state components $\psi_{n,mA}$ and $\psi_{n,mB}$ of $|\psi\rangle$ for two sublattice sites A and B at the cell spatial coordinate ($n,m$) are defined by
\begin{equation}
\psi_{n,mA}=w^n\sin(n\phi_j)r^{m-1}\mathcal{T}(m)\sin(\theta_\ell), \label{Eq13}
\end{equation}
\begin{equation}
\psi_{n,mB}=(E-2Q)w^n\sin(n\phi_j)r^m\sin(m\theta_\ell), \label{Eq14}
\end{equation}
where $r=\sqrt{t_{1R}/t_{1L}}$ and $w=\sqrt{g_{1U}/g_{1D}}$. We find that the generalized Brillouin zones (GBZs) of the bulk states can be exactly defined by $\beta_x=r\exp(\pm i\theta_\ell)$ and $\beta_y=w\exp(\pm i\phi_j)$ and thus the NHSE occurs whenever $r\neq1$ or $w\neq1$. When the parameters satisfy $|t_2/\sqrt{t_{1L}t_{1R}}|>1$ \cite{Hou23}, the topological edge states will pop up, but with different edge energies $E^{\mathrm{dOBC}}_\mathrm{edge}=2\mu Q$ or $2Q$.

\begin{figure}
	[ht!]
	\begin{center}
		\includegraphics[width=8.6cm]%
		{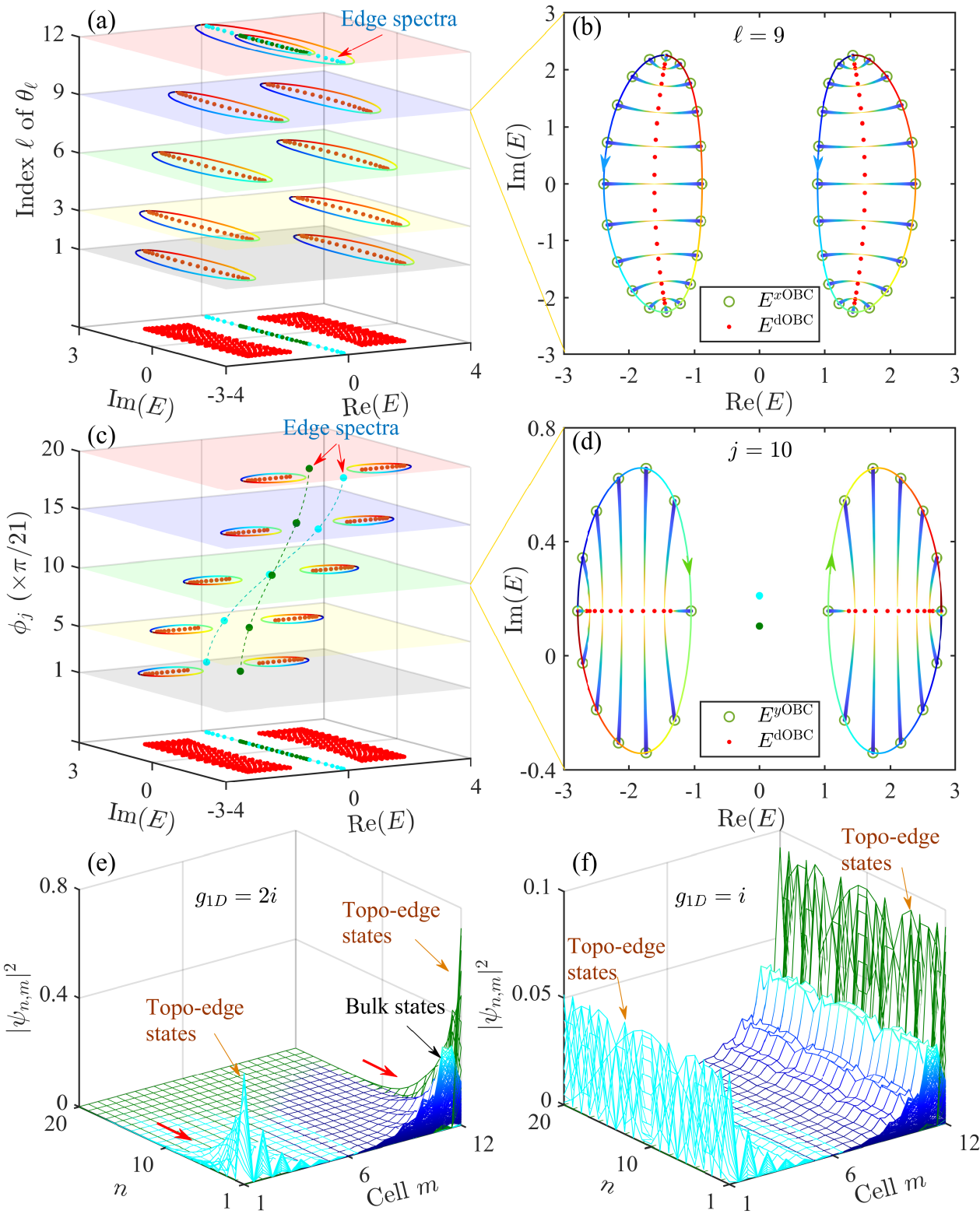}%%
		\caption{Demonstration of skin-topological edge states in a 2D SSH-HN lattice with dOBC ($M=12$, $N=20$), based on analytical solutions (\ref{Eq12})--(\ref{Eq14}) using $t_{1L}= 1/3$, $t_{1R}=4/3$, $t_2=2$, $g_{1D}=2i$, $g_{1U}=i$, and $\mu=1/2$. (a,b) Energy spectra $E^{\mathrm{dOBC}}$ [red dots, defined by Eq. (\ref{Eq12})] surrounded by the $E^{x\mathrm{dOBC}}$ spectra [colored directional curves, Eq.~(\ref{Eq7})] for given index $\ell$; (c,d) Energy spectra $E^{\mathrm{dOBC}}$ surrounded by the $E^{y\mathrm{dOBC}}$ spectra [Eq.~(\ref{Eq10})] for given $\phi_j=j\pi/(N+1)$; (e) The normalized bulk corner states and skin-topological edge states. The bold colored lines in (b,d) indicate the energy flow $E^{x\mathrm{OBC}}\rightarrow E^{\mathrm{dOBC}}$, where $\varphi=0$, $\vartheta=0\rightarrow-7i$, or $E^{y\mathrm{OBC}}\rightarrow E^{\mathrm{dOBC}}$, where $\vartheta=0$, $\varphi=0\rightarrow8i$. We also demonstrate in (f) the skin-effect-free topological edge states, by only changing the value of $g_{1D}$ from $2i$ to $i$. These analytical solutions are confirmed to be entirely consistent with numerical ones (see Supplementary Fig. 2 in SM).} \label{fig2}
	\end{center}
\end{figure}

From the exact solutions (\ref{Eq12})--(\ref{Eq14}), we find that the NHSE can occur for both the bulk and topological edge states in a fully open 2D lattice. Where exactly these NHSE-pumped bulk or edge modes tend to accumulate can be determined by the winding direction of $E^{x\mathrm{OBC}}$ (or $E^{y\mathrm{OBC}}$) with respect to $E^{\mathrm{dOBC}}$, as momentum $k_y=2j\pi/N$ (or $k_x=2\ell\pi/M$) runs along its first BZ. In principle, the clockwise (counter-clockwise) winding implies that the bulk or edge states accumulate upwards (downwards) in the $y$ direction, or move rightwards (leftwards) in the $x$ direction. For illustration, we demonstrate in Fig. \ref{fig2} that $E^{x\mathrm{OBC}}$ rotates counter-clockwise around $E^{\mathrm{dOBC}}$ (red dots) for given $\theta_\ell$ [see Figs. \ref{fig2}(a) and \ref{fig2}(b)], while $E^{y\mathrm{OBC}}$ surrounds the $E^{\mathrm{dOBC}}$ clockwise for given $\phi_j$ [see Figs. \ref{fig2}(c) and \ref{fig2}(d)], resulting in the bulk corner states located on the lower right corner [see Fig. \ref{fig2}(e)]. Meanwhile, two topological edge states on the opposite edges [see green and cyan lines in Fig. \ref{fig2}(e)] arise for $\mu=1/2$ (inhomogeneity), displaying non-degenerate eigenenergies $2\mu Q$ (green circles) and $2Q$ (cyan circles) as seen in Fig. \ref{fig2}(c) \cite{Remark1}. These gapped edge states undergo a strong NHSE and accumulate downwards along the edges, as indicated by red arrows in Fig. \ref{fig2}(e). If only the parameter $g_{1D}$ is modified as $g_{1D}=i$, which implies $w=1$, then they are extensively distributed along the edges, although the bulk states still exhibit skin  effect [see Fig. \ref{fig2}(f)]. Since these topological edge states enjoy a point-gap topology (i.e., skin effect), we term them skin-topological edge states as proposed in Ref.~\cite{Lee19}. From the above observation, we now propose two simple yet universal winding numbers for predicting these NHSEs in 2D:
\begin{equation}
W_x=\oint_0^{2\pi}\frac{dk_x}{2\pi i} \partial_{k_x} \ln\{\mathrm{det}[H(k_x,k_y=\phi-i\ln w)-E_{\mathrm{OBC}}]\}, \label{Eq15}
\end{equation}
\begin{equation}
W_y=\oint_0^{2\pi}\frac{dk_y}{2\pi i} \partial_{k_y} \ln\{\mathrm{det}[H(k_x=\theta-i\ln r,k_y)-E_{\mathrm{OBC}}]\}, \label{Eq16}
\end{equation}
where $\phi,\theta \in[0,\pi]$ (half BZ) for the bulk states \cite{Remark2} but $\theta=\pi-\arccos[(\sqrt{t_{1L}t_{1R}}/t_2+t_2/\sqrt{t_{1L}t_{1R}})/2]$ for the topological edge states. For the current parameters used, the complex integrations (\ref{Eq15}) and (\ref{Eq16}) yield exactly $W_x^{\rm bulk}=1$, $W_x^{\rm edge}=0$, $W_y^{\rm bulk}=W_y^{\rm edge}=-1$ \cite{Needham23}, implying that the bulk states accumulate towards the positive $x$ and negative $y$ directions, but the topological edge ones only aggregate along the $-y$ direction, completely consistent with our analytical solutions shown in Figs. \ref{fig2}(a)--\ref{fig2}(e).

Subsequently, we consider the more complicated SSH-SSH rectangular lattice shown in Fig. \ref{fig1}(b) \cite{Lee19,LiuT19,LiuF17}. The Hamiltonian of such 2D lattice reads
\begin{eqnarray}
\hat{H}^{\mathrm{SS}}&=&\sum_{n,m}(\hat{C}^{\dagger}_{n,m}M_1\hat{C}_{n,m}+\hat{C}^{\dagger}_{n,m+1}M_{2}^\dag\hat{C}_{n,m}
+\hat{C}^{\dagger}_{n,m}M_{2}\hat{C}_{n,m+1}\nonumber\\
&&+\hat{C}^{\dagger}_{n+1,m}M_{3}^\dag\hat{C}_{n,m}+\hat{C}^{\dagger}_{n,m}M_{3}\hat{C}_{n+1,m})
+\hat{H}_{\mathrm{B3}}+\hat{H}_{\mathrm{B4}}, \label{Eq17}
\end{eqnarray}
where $\hat{C}^{\dagger}_{n,m} = (\hat{a}^{\dagger}_{n,m}, \hat{b}^{\dagger}_{n,m},\hat{c}^{\dagger}_{n,m}, \hat{d}^{\dagger}_{n,m})$ are the creation operators of particles on sublattices A, B, C, and D at the cell coordinate $(n,~m)$ $(n=1,\cdots,N;~m=1,\cdots,M)$, and
\begin{eqnarray}
	M_{1}&=&	\left[\begin{array}{cccc}
		0 &  t_{1L} &  g_{1D} &  0 \\
		t_{1R} &0   & 0  & g_{3D}    \\
        g_{1U} &0   & 0 &   t_{3L}   \\
        0 & g_{3U}  &   t_{3R}   & 0 \\		
	\end{array}\right],\nonumber\\
M_{2}&=&	\left[\begin{array}{cccc}
		0 &  0 &    0  &  0 \\
		t_{2} &0  & 0  &   0     \\
0&0   & 0 &   0   \\
0   &  0  &   t_{4}   & 0 \\		
	\end{array}\right],~~~
M_{3}=	\left[\begin{array}{cccc}
		0 &  0 &    0  &  0 \\
		0 &0  & 0&    0    \\
g_{2}  &0   & 0 &   0   \\
0 & g_{4}   &   0   & 0 \\		
	\end{array}\right], \label{Eq18}
\end{eqnarray}
with $t$'s and $g$'s being the real hopping parameters.
$\hat{H}_{\mathrm{B3}}=\sum_{n}(\delta_1\hat{C}^{\dagger}_{n,1}M_{2}^{\dagger}\hat{C}_{n,M}
+\delta_2\hat{C}^{\dagger}_{n,M}M_{2}\hat{C}_{n,1})$ and $\hat{H}_{\mathrm{B4}}=\sum_{m}(\kappa_1\hat{C}^{\dagger}_{1,m}M_{3}^{\dagger}\hat{C}_{N,m}
+\kappa_2\hat{C}^{\dagger}_{N,m}M_{3}\hat{C}_{1,m})$ are the boundary conditions and one can classify them into dGBC ($\delta_1=\delta_2^{-1}=e^{i\varphi}$, $\kappa_1=\kappa_2^{-1}=e^{i\vartheta}$), $x$OBC ($\hat{H}_{\mathrm{B3}}=0$, $\kappa_1=\kappa_2^{-1}=e^{i\vartheta}$), $y$OBC ($\delta_1=\delta_2^{-1}=e^{i\varphi}$, $\hat{H}_{\mathrm{B4}}=0$), and dOBC ($\hat{H}_{\mathrm{B3}}=\hat{H}_{\mathrm{B4}}=0$), respectively. We note that this 2D Hamiltonian respects the sublattice symmetry \cite{Kawa19,Li24} and thus its eigenenergies will come in pairs ($E,~-E$). We should point out that our Hamiltonian (\ref{Eq17}) has significantly generalized those adopted in Refs. \cite{LiuT19,LiuF17}.

In a similar fashion, one can exactly solve the eigenvalue equation of Hamiltonian (\ref{Eq17}) in real space for the above boundary conditions, if the hopping parameters fulfil $t_{3Z}/t_{1Z}=t_4/t_2=\mu$ and $g_{3X}/g_{1X}=g_4/g_2=\nu$, where $Z=L,R$ and $X=U,D$ (see Sec. II in SM). Specifically, under dGBC, the energy spectrum of Hamiltonian (\ref{Eq17}) can be written as
\begin{equation}
E^{\mathrm{dGBC}}=\pm\sqrt{\frac{(\mu^2+1)T+(\nu^2+1)G\pm\sqrt{\Delta(T,G)}}{2}}, \label{Eq19}
\end{equation}
where $T$ is given by Eq. (\ref{Eq6}), and
\begin{equation}
G=(g_{1U}+g_{2}\omega_j)\left(g_{1D}+\frac{g_{2}}{\omega_j}\right), \label{Eq20}
\end{equation}
\begin{equation}
\Delta(T,G)=\left[(\mu+1)^2T+(\nu-1)^2G\right]\left[(\mu-1)^2T+(\nu+1)^2G\right], \label{Eq21}
\end{equation}
with $\varpi_\ell$ and $\omega_j$ being exactly the same as in Eq. (\ref{Eq6}). Therefore, replacing $T$ by $R$ in Eq.~(\ref{Eq19}), where $R$ is defined by Eq. (\ref{Eq8}), one obtains the energy spectrum under $x$OBC,
\begin{equation}
E^{x\mathrm{OBC}}=\pm\sqrt{\frac{(\mu^2+1)R+(\nu^2+1)G\pm\sqrt{\Delta(R,G)}}{2}}. \label{Eq22}
\end{equation}
Similarly, if replacing $G$ by $S$ in Eq. (\ref{Eq19}), we arrive at the energy spectrum under $y$OBC:
\begin{equation}
E^{y\mathrm{OBC}}=\pm\sqrt{\frac{(\mu^2+1)T+(\nu^2+1)S\pm\sqrt{\Delta(T,S)}}{2}}, \label{Eq23}
\end{equation}
where
\begin{equation}
S=2 g_{2} \sqrt{g_{1 D} g_{1 U}}\cos(\phi_j)+g_{1 D} g_{1 U}+g_{2}^{2}. \label{Eq24}
\end{equation}
The $\phi_j$ in Eq. (\ref{Eq24}) is one of $N$ complex roots of the equation $\mathcal{G}(N+1)=0$, where
\begin{equation}
\mathcal{G}(n)=\frac{g_{2}\sin[(n-1)\phi_j]}{\sin(\phi_j)}+\frac{\sqrt{g_{1 D} g_{1 U}}\sin(n\phi_j)}{\sin(\phi_j)}. \label{Eq25}
\end{equation}

What we are primarily concerned with are the solutions with dOBC, which possess the following energy spectra:
\begin{equation}
E^{\mathrm{dOBC}}=\pm\sqrt{\frac{(\mu^2+1)R+(\nu^2+1)S\pm\sqrt{\Delta(R,S)}}{2}}.\label{Eq26}
\end{equation}
Correspondingly, the state components of $|\psi\rangle$ at the cell spatial coordinate $(n,~m)$ are given by
\begin{equation}
\psi_{n,mA}=w^{n-1}\mathcal{G}(n)\sin(\phi_j)r^{m-1}\mathcal{T}(m)\sin(\theta_\ell), \label{Eq27}
\end{equation}
\begin{equation}
\psi_{n,mB}=E_{B}w^{n-1}\mathcal{G}(n)\sin(\phi_j)r^m\sin(m\theta_\ell), \label{Eq28}
\end{equation}
\begin{equation}
\psi_{n,mC}=E_{C}w^n\sin(n\phi_j)r^{m-1}\mathcal{T}(m)\sin(\theta_\ell), \label{Eq29}
\end{equation}
\begin{equation}
\psi_{n,mD}=E_{D}w^n\sin(n\phi_j)r^m\sin(m\theta_\ell), \label{Eq30}
\end{equation}
where $\mathcal{T}(m)$ is defined by Eq.~(\ref{Eq9}), $\mathcal{G}(n)$ is given by Eq.~(\ref{Eq25}), and $E_{B}=(\mu\nu+1)R E/\Gamma$, $E_{C}=(E^2-R-\nu^2S)E/\Gamma$, and $E_{D}=(\mu E^2-\mu R+\nu S)R/\Gamma$, with $\Gamma=E^2+\mu\nu R-\nu^2S$. Under such dOBC, the GBZs are found to be $\beta_x=r\exp(\pm i\theta_\ell)$ and $\beta_y=w\exp(\pm i\phi_j)$, which trace a circle in the complex plane (see Sec. III in SM).

\begin{figure}
	[ht!]
	\begin{center}
		\includegraphics[width=8.6cm]%
		{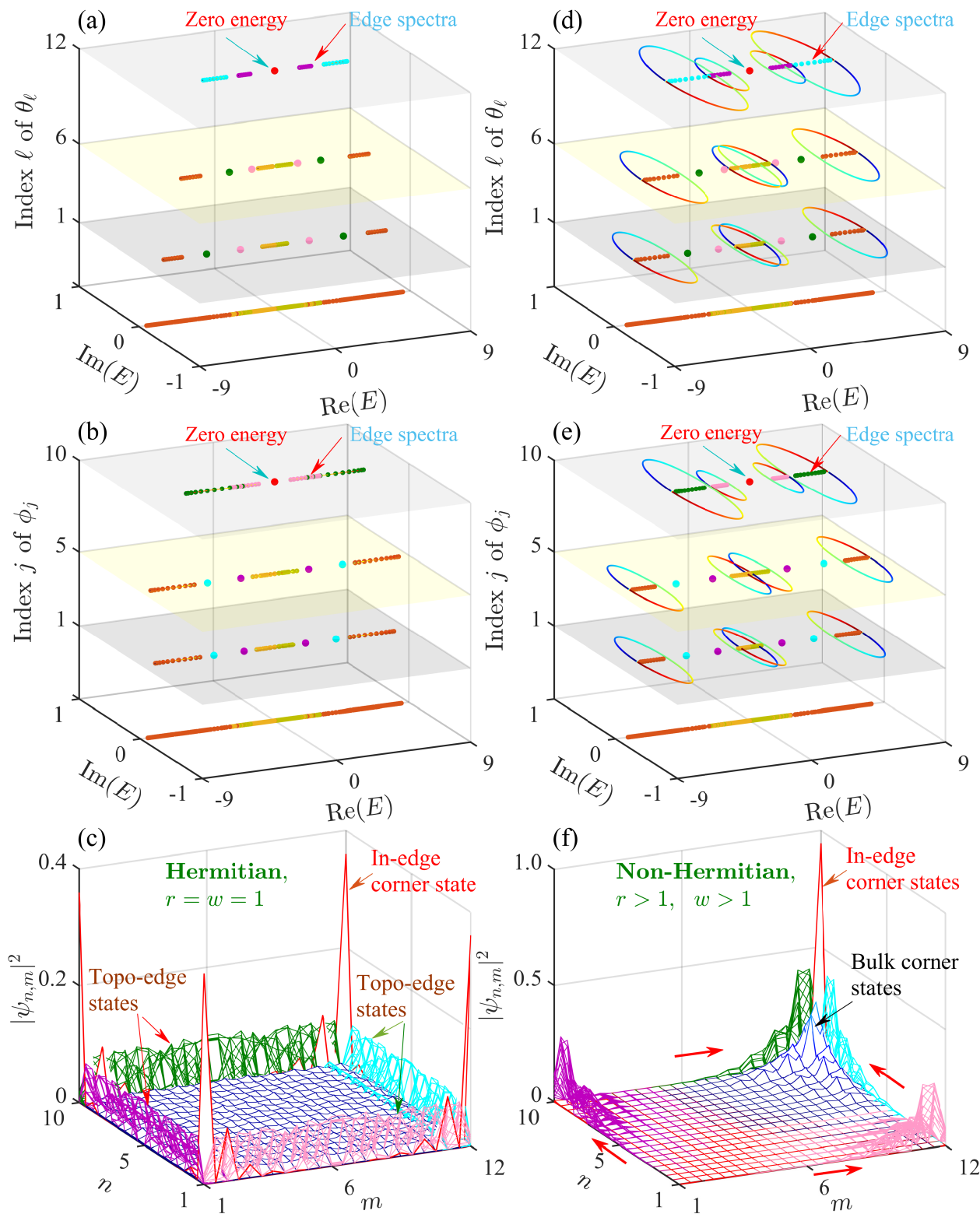}%%
		\caption{Enhanced in-edge corner states arising from the interplay between topological edge states and NHSE in a 2D SSH-SSH lattice with dOBC ($M=12$, $N=10$), calculated from analytical solutions (\ref{Eq26})--(\ref{Eq30}). (a--c) show the Hermitian case ($t_{1L}= 1$, $g_{1U}=2/5$), where gapped edge states (cyan, purple, green, and pink lines) stand extensively on four sides and four topological corner states (red lines) form on different corners. (d--f) show the non-Hermitian case ($t_{1L}= 1/4$, $g_{1U}=6/5$ ), where gapped edge states undergo NHSE and the in-edge corner states build on only one corner. The other parameters are given by $t_{1R}=1$, $g_{1D}=2/5$, $t_2=g_2=\mu=\nu=2$. (a,d) Energy spectra $E^{\mathrm{dOBC}}$ [Eq.~(\ref{Eq26})] versus $\ell$, (b,e) energy spectra $E^{\mathrm{dOBC}}$ versus $j$, and (c,f) the normalized bulk, edge, and in-edge corner states,  which agree well with numerical results (see Supplementary Fig. 5 in SM).} \label{fig3}
	\end{center}
\end{figure}

An inspection of the above closed-form solutions (\ref{Eq27})--(\ref{Eq30}) reveals that the topological edge states could emerge on both the left and right sides for $|t_2/\sqrt{t_{1L}t_{1R}}|>1$ \cite{Hou23} and $\nu\neq1$, with eigenenergies $E^{\mathrm{dOBC}}_\mathrm{edge}=\pm\nu\sqrt{S}$ or $\pm\sqrt{S}$ (see Supplementary Fig. 3 in SM). If further $|g_2/\sqrt{g_{1 D} g_{1 U}}|>1$ is met, along with $\mu\neq1$, the gapped edge states could emerge on the four sides of 2D lattice, with eigenenergies $E^{\mathrm{dOBC}}_\mathrm{edge}=\pm\nu\sqrt{S}$ (cyan circles), $\pm\sqrt{S}$ (purple circles), $\pm\mu\sqrt{R}$ (green circles), or $\pm\sqrt{R}$ (pink circles), as exhibited in Fig. \ref{fig3}. In the Hermitian case [see Figs. \ref{fig3}(a)--\ref{fig3}(c)], these gapped edge states distribute extensively on the four edges because of $W_x^{\rm edge}=W_y^{\rm edge}=0$ calculated from Eqs. (\ref{Eq15}) and (\ref{Eq16}). However, in the non-Hermitian case [see Figs. \ref{fig3}(d)--\ref{fig3}(f)], the winding numbers become $W_x^{\rm edge}=W_y^{\rm edge}=1$, and thus the topological edge states undergo NHSE along the edges, as indicated by red arrows in Fig. \ref{fig3}(f). More intriguingly, different from the SSH-HN lattice discussed earlier, the 2D SSH-SSH lattice could admit the emergence of zero-energy topological corner states [i.e., $E^{\mathrm{dOBC}}_{\mathrm{corner}}=0$, indicated by red circles in Figs. \ref{fig3}(a,b,d,e)], which occur on the corner where the gapped edge states on adjacent edges intersect [see red lines in Figs. \ref{fig3}(c) and \ref{fig3}(f)]. In the former Hermitian case, there appear four in-edge corner states, each having an intensity around $(1-t_{1L}t_{1R}/t_2^2)(1-g_{1D}g_{1U}/g_2^2)/2=0.36$. However, in the non-Hermitian case, as the gapped edge states redistribute on the edges because of NHSE, the in-edge corner states build on only one corner, with intensity around $0.94$, more than twice that seen in the Hermitian case. In fact, though they correspond to a diabolic point in Hermitian case, these in-edge corner states manifest as an exceptional point in the non-Hermitian case \cite{Ozd19,Lee22,Liu2024}.

In conclusion, we obtained the exact closed-form solutions for both skin modes and topological edge states in two typical 2D non-Hermitian lattices, under different boundary conditions. We confirmed analytically the higher-order topology of gapped edge states \cite{Lee19,Zou21,Zhang21}, which may undergo NHSE along the edges. The explicit energy-spectrum relationships established allow us to propose two simple yet universal topological winding numbers to predict accurately the NHSE occurring for both the bulk and topological edge states. For the non-Hermitian SSH-SSH lattice with dOBC, we also discovered that there would appear a strongly localized in-edge corner state, which arises from the interplay between gapped edge states and NHSE, with intensity neither accessible to the Hermitian counterpart nor to the bulk corner states \cite{Shang22,Hou24}. We expect that these unusual topological states not only enrich the non-Bloch band topology \cite{Yok19,Shen18,Hou22,Yang24} in two and higher dimensions but also trigger experimental investigations into many related phenomena such as edge topological lasing \cite{Harari18,Bandres18,Lu2024}.

%\noindent
\begin{acknowledgments}
This work was supported by the National Natural Science Foundation of China (Grants No. 12374301 and No. 11974075) and the National Key Research and Development Program of China (Grant No. 2021YFA1200700). C. H. L acknowledges support
from the National Research Foundation, Singapore under its QEP2.0 programme (NRF2021-QEP2-02-P09), and the Ministry of Education, Singapore (MOE Award No. MOE-T2EP50222-0003).
\end{acknowledgments}

%\bibliographystyle{myprl}
%\bibliography{sampleBibFile}

\begin{thebibliography}{99}
\bibitem{Remark2} The result $\phi,\theta \in[0,\pi]$, instead of $\in[0,2\pi]$, stems from our analytical solutions (\ref{Eq7}), (\ref{Eq10}) and (\ref{Eq12}). In the thermodynamic limit ($M\rightarrow\infty,~~N\rightarrow\infty$), one can see from Eq. (\ref{Eq11}) that the value of $\phi$ will run continuously from $0$ to $\pi$ and from Eq. (\ref{Eq9}) that the root $\theta$ of $\mathcal{T}(M+1)=0$ also runs continuously from $0$ to $\pi$, when only the bulk states are considered. This makes our winding numbers defined here be distinctly different from Eq. (34) in Ref. \cite{Kawabata20} that fails to predict the non-Hermitian skin effect in 2D.
\end{thebibliography}

\end{document}

\title{Supplementary Material for ``Exact Solutions Disentangle Higher-Order Topology in 2D Non-Hermitian Lattices''}

\author{Lingfang Li}
\affiliation{Key Laboratory of Quantum Materials and Devices of Ministry of Education, School of Physics, Frontiers Science Center for Mobile Information Communication and Security, Southeast University, Nanjing 211189, China}

\author{Yating Wei}
\affiliation{Key Laboratory of Quantum Materials and Devices of Ministry of Education, School of Physics, Frontiers Science Center for Mobile Information Communication and Security, Southeast University, Nanjing 211189, China}

\author{Gangzhou Wu}
\affiliation{Key Laboratory of Quantum Materials and Devices of Ministry of Education, School of Physics, Frontiers Science Center for Mobile Information Communication and Security, Southeast University, Nanjing 211189, China}

\author{Yang Ruan}
\affiliation{Key Laboratory of Quantum Materials and Devices of Ministry of Education, School of Physics, Frontiers Science Center for Mobile Information Communication and Security, Southeast University, Nanjing 211189, China}

\author{Shihua Chen}
%\email{cshua@seu.edu.cn}
\affiliation{Key Laboratory of Quantum Materials and Devices of Ministry of Education, School of Physics, Frontiers Science Center for Mobile Information Communication and Security, Southeast University, Nanjing 211189, China}
\affiliation{Purple Mountain Laboratories, Nanjing 211111, China}

\author{Ching Hua Lee}
%\email{phylch@nus.edu.sg}
\affiliation{Department of Physics, National University of Singapore, Singapore 117551, Republic of Singapore}

\author{Zhenhua Ni}
%\email{zhni@seu.edu.cn}
\affiliation{Key Laboratory of Quantum Materials and Devices of Ministry of Education, School of Physics, Frontiers Science Center for Mobile Information Communication and Security, Southeast University, Nanjing 211189, China}
\affiliation{Purple Mountain Laboratories, Nanjing 211111, China}

%\date{\today}
\maketitle

     \vspace{1.5cm}

The supplementary material is organized as follows. In Secs. \ref{Sec1} and \ref{Sec2}, we solve the 2D SSH-HN lattice model and the 2D SSH-SSH lattice model in real space, respectively, using three different boundary conditions. The derivations of the generalized Brillouin zone (GBZ) and the parameter conditions for topological edge states are presented in Sec. \ref{Sec3}.

\section{The 2D SSH-HN lattice model} \label{Sec1}

In this section, we provide the details of derivation of exact solutions of the 2D SSH-SH lattice model that governs the hopping dynamics in rectangular lattice illustrated in Fig. 1(a) in the main text \cite{Edva2022}, under three different types of boundary conditions given by $\delta_{1,2}$ and $\kappa_{1,2}$ in the two directions. The real-space Hamiltonian of such 2D model reads
\begin{align}
\hat{H}^{\mathrm{SN}}=&\sum_{n,m}(\hat{C}^{\dagger}_{n,m}M_1\hat{C}_{n,m}+\hat{C}^{\dagger}_{n,m+1}M_{2}^{\dagger}
\hat{C}_{n,m}+\hat{C}^{\dagger}_{n,m}M_{2}\hat{C}_{n,m+1}+\hat{C}^{\dagger}_{n+1,m}M_{U}\hat{C}_{n,m}
+\hat{C}^{\dagger}_{n,m}M_{D}\hat{C}_{n+1,m})  \nonumber \\
&+\sum_{n}(\delta_1\hat{C}^{\dagger}_{n,1}M_{2}^{\dagger}
\hat{C}_{n,M}+\delta_2\hat{C}^{\dagger}_{n,M}M_{2}\hat{C}_{n,1})\nonumber
+\sum_{m}(\kappa_1\hat{C}^{\dagger}_{1,m}M_{U}\hat{C}_{N,m}+\kappa_2\hat{C}^{\dagger}_{N,m}M_{D}\hat{C}_{1,m}), \tag{S1}\label{S1}
\end{align}
where $\hat{C}^{\dagger}_{n,m} = (\hat{a}^{\dagger}_{n,m}, \hat{b}^{\dagger}_{n,m})$ are the creation operators of particles on sublattices A and B in the cell spatial coordinate ($n,m$), that is, in the $n$th row and $m$th column of the 2D lattice, and
\begin{align}
M_1=	\left[\begin{array}{cc}
		0 &  t_{1L}  \\
		t_{1R} &0    \\		
	\end{array}\right],~~~~
M_{2}=	\left[\begin{array}{cc}
		0 &  0 \\
		t_{2} &  0  \\
	\end{array}\right],~~~~
M_{X}=	\left[\begin{array}{cc}
		g_{1X} &  0  \\
		0 &  g_{3X}   \\	
	 \end{array}\right],~~~(X=U, D). \tag{S2}\label{S2}
\end{align}
Here, the system parameters $t$'s and $g$'s represent the hopping amplitudes.

First, expressed in an appropriate tensor product basis in 2D real space, we write this Hamiltonian into a $2MN\times2MN$ matrix:
\begin{align}
	H^{\mathrm{SN}}=	\left[\begin{array}{ccccc}
		K &K_{D}  &             &     & \kappa_{1} K_{U}       \\
		K_{U} &K  & K_{D} &     &              \\
		          & K_{U}  & K & \ddots &      \\
		          &              &   \ddots     &\ddots &K_{D}        \\
		\kappa_{2} K_{D}  &           &      &K_{U}  &K  \\		
	\end{array}\right]_{2MN\times2MN}. \tag{S3}\label{S3}
\end{align}
with $\kappa_{1} K_{U}$ and $\kappa_{2} K_{D}$ at two corners denoting the boundary conditions along the $y$ direction. Here $K$ and $K_{X}$ $(X=U, D)$ are a $2M\times2M$  matrix composed of hopping parameters, defined by
\begin{align}
	K=	\left[\begin{array}{cccccc}
0         & t_{1L}  & & & &  \delta_1 t_{2}  \\
t_{1R} &0          &t_{2} & &    & \\
                 &t_{2}   &0         &t_{1L} & & \\
	&        &  t_{1R}  & 0 & \ddots  &   \\
	&         & &               \ddots     &\ddots &t_{1L}        \\
	\delta_2 t_{2}  &   &     &        &t_{1R}  &0\\		
	\end{array}\right]_{2M\times2M},~~~~
	K_{X}=	\left[\begin{array}{cccccc}
		 g_{1X} &   &             &  &   &        \\
		 & g_{3X} &  &     &       &       \\
		          &   & g_{1X} &   &   &   \\
                  & &   & g_{3X} &   &      \\
		          &              &   &     &\ddots &     \\
		  &           &  &    & &g_{3X}  \\		
	\end{array}\right]_{2M\times2M},  \tag{S4}\label{S4}
\end{align}
with $\delta_{1,2}$ denoting the boundary conditions along the $x$ direction.
Meanwhile, the eigenvector $|\Psi^{\mathrm{SN}}\rangle$ of Hamiltonian (\ref{S1}) can be arranged into a column array:
\begin{align}
|\Psi^{\mathrm{SN}}\rangle & = \left[\begin{array}{c}
|\psi_{1}\rangle \\
|\psi_{2}\rangle\\
|\psi_{3}\rangle \\
\vdots \\
|\psi_{N}\rangle
\end{array}\right],  \tag{S5}\label{S5}
\end{align}
where
\begin{align}
|\psi_{n}\rangle=[\underbrace{\psi_{n,1A},\psi_{n,1B},\psi_{n,2A},\psi_{n,2B},\cdots,\psi_{n,mA},
\psi_{n,mB},\cdots,\psi_{n,MA},\psi_{n,MB}}_{2M}]^{\mathrm{T}},~~~(n=1,\cdots, N), \tag{S6}\label{S6}
\end{align}
denotes the state components on the $n$th row of the lattice (here the superscript $\mathrm{T}$ means transpose operation).

Now, with the above matrix forms, the eigenvalue equation of Hamiltonian (\ref{S1}):
\begin{align}
H^{\mathrm{SN}}|\Psi^{\mathrm{SN}}\rangle=E|\Psi^{\mathrm{SN}}\rangle, \tag{S7}\label{S7}
\end{align}
can be transformed into a system of bulk equations, expressed as
\begin{align}
K|\psi_{n}\rangle+K_{D}|\psi_{n+1}\rangle+K_{U}|\psi_{n-1}\rangle=E|\psi_{n}\rangle,  \tag{S8}\label{S8}
\end{align}
where $n=2,\cdots,N-1$, and two boundary equations given by
\begin{align}
K|\psi_{1}\rangle+K_{D}|\psi_{2}\rangle+\kappa_1 K_{U}|\psi_{N}\rangle=&E|\psi_{1}\rangle, \tag{S9}\label{S9}
\end{align}
\begin{align}
K|\psi_{N}\rangle+\kappa_2 K_{D}|\psi_{1}\rangle+K_{U}|\psi_{N-1}\rangle=&E|\psi_{N}\rangle.  \tag{S10}\label{S10}
\end{align}
In the following, we can solve Eqs. (\ref{S8})--(\ref{S10}) readily, using different boundary conditions denoted by $\delta_{1,2}$ and $\kappa_{1,2}$. Of course, in order for these equations to allow exact closed-form solutions for all boundary conditions, we will assume $g_{3X}=\mu g_{1X}$ ($\mu\in \mathds{R}$), unless otherwise mentioned.

Before proceeding, let us write the momentum-space Hamiltonian for such 2D SSH-HN lattice as
\begin{align}
	H^{\mathrm{SN}}(k_{x}, k_{y})=	\left[\begin{array}{cc}
		 g_{1D}\exp(ik_{y})+g_{1U}\exp(-ik_{y}) & t_{1L}+t_{2}\exp(-ik_{x}) \\
		t_{1R}+t_{2}\exp(ik_{x})   & g_{3D}\exp(ik_{y}) + g_{3U}\exp(-ik_{y}) \\
	\end{array}\right], \tag{S11}\label{S11}
\end{align}
which follows by performing Fourier transformation
\begin{align}
\hat{a}_{n,m}=\frac{1}{\sqrt{M N}} \sum_{k_{x}, k_{y}} e^{i k_{x} m+i k_{y} n} \hat{a}_{k_{x}, k_{y}},~~~~\hat{b}_{n,m}=\frac{1}{\sqrt{M N}} \sum_{k_{x}, k_{y}} e^{i k_{x} m+i k_{y} n} \hat{b}_{k_{x}, k_{y}},  \tag{S12}\label{S12}
\end{align}
on real-space Hamiltonian (\ref{S1}) based on the assumption of the translational invariance in the bulk \cite{Shen13}. This Bloch Hamiltonian can always be used to determine the band structure of the bulk, by solving the characteristic equation det$[E-H^{\mathrm{SN}}(k_{x}, k_{y})] =0$. Besides, it is easy to show that $\sigma_z\mathcal{H}^{\mathrm{SN}*}(k_{x}, k_{y})\sigma_z^{-1}=-\mathcal{H}^\mathrm{SN}(-k_{x}, -k_{y})$ for $g_{1,3X}\in i\mathds{R}$ (dissipative rates) or  $\mathcal{H}^{\mathrm{SN}*}(k_{x}, k_{y})=\mathcal{H}^\mathrm{SN}(-k_{x}, -k_{y})$ for $g_{1,3X}\in \mathds{R}$ (hopping rates) (here $\sigma_z$ is the Pauli spin matrix and the asterisk means complex conjugation). Therefore, one can conclude that the Hamiltonian (\ref{S1}) respects the time-reversal symmetry (TRS) and hence the eigenenergies will come in $(E, -E^*)$ pairs in the former case but in $(E, E^*)$ pairs in the latter case \cite{Kawa19,Li24}.

\subsection{Exact solutions under double generalized periodic boundary conditions }\label{Sec1A}

Firstly, we consider the simplest double generalized periodic boundary condition (dGBC), which implies $\delta_1=\delta_2^{-1}=e^{i\varphi}$ and $\kappa_1=\kappa_2^{-1}=e^{i\vartheta}$. Noting that, when $\vartheta=\varphi=0$, this dGBC reduces to the familiar double periodic boundary condition (dPBC), which means that the 2D lattice is periodic along both the $x$ and $y$ directions. Under the circumstances, solving Eqs. (\ref{S8})--(\ref{S10}), one can obtain readily the exact eigenstate solutions $|\Psi^{\mathrm{SN}}\rangle$, with the following state components
\begin{align}
\psi_{n,mA}=\omega_j^{n-1}\varpi_\ell^m,~~~~\psi_{n,mB}=\frac{E-P}{t_{1L}+t_{2}/\varpi_\ell}\omega_j^{n-1}\varpi_\ell^m
=\frac{E-P}{t_{1L}+t_{2}/\varpi_\ell}\psi_{n,mA}, \tag{S13}\label{S13}
\end{align}
along with the associated eigenenergy $E$ given by
\begin{align}
E=&\frac{(\mu+1) P \pm\sqrt{(\mu-1)^2P^2+4T}}{2}\equiv E^{\mathrm{dGBC}}, \tag{S14}\label{S14}
\end{align}
where
\begin{align}
P=g_{1D} \omega_j+\frac{g_{1U}}{\omega_j},~~~T=(t_{1R}+t_{2}\varpi_\ell)(t_{1L}+\frac{t_{2}}{\varpi_\ell}), \tag{S15}\label{S15}
\end{align}
\begin{align}
\varpi_\ell=\exp\left(\frac{i2\ell\pi}{M}-\frac{i\varphi}{M}\right),~~~
\omega_j=\exp\left(\frac{i2j\pi}{N}-\frac{i\vartheta}{N}\right). \tag{S16}\label{S16}
\end{align}
It should be noted that the subscripts $n$ and $m$ in Eq. (\ref{S13}) denote the cell spatial coordinate, while the subscripts $j=1,\cdots,N$ and $\ell=1,\cdots,M$ are used for labelling the eigenstates, which will have a total number of $2M\times N$.

Of course, one can also derive the eigenenergy (\ref{S14}) directly from the characteristic equation of the Bloch Hamiltonian (\ref{S11}). But the merit of our solutions is that they provide not only the energy spectrum but also the real-space state distributions. It is clear that, under dGBC, there will be no non-Hermitian skin effects even if the system under study is non-Hermitian.

\subsection{Exact solutions under unidirectionally open boundary condition}\label{Sec1B}

For the sake of discussion, let us classify the single-direction open boundary condition (OBC) further into two categories: one is $x$OBC, corresponding to $\kappa_1=\kappa_2^{-1}=e^{i\vartheta}$ but $\delta_{1,2}=0$, which means that the 2D lattice is open along the $x$ direction, but periodic along the $y$ direction, while the other is $y$OBC, defined by $\delta_1=\delta_2^{-1}=e^{i\varphi}$ and $\kappa_{1,2}=0$, implying that the 2D lattice is open along the $y$ direction, but periodic along the $x$ direction.

Under $x$OBC, one can again solve Eqs. (\ref{S8})--(\ref{S10}) exactly, with the following closed-form solutions
\begin{align}
\psi_{n,mA}=\omega_j^{n-1}r^{m-1}\mathcal{T}(m)\sin(\theta_\ell),~~~~\psi_{n,mB}=(E-P)\omega_j^{n-1}r^m\sin(m\theta_\ell),   \tag{S17}\label{S17}
\end{align}
\begin{align}
E=\frac{(\mu+1) P \pm\sqrt{(\mu-1)^2 P^2+4R}}{2}\equiv E^{x\mathrm{OBC}}, \tag{S18}\label{S18}
\end{align}
where
\begin{align}
R=2t_{2}\sqrt{t_{1L}t_{1R}}\cos(\theta_\ell)+t_{1L}t_{1R}+t_{2}^2,~~~~ r=\sqrt{\frac{t_{1R}}{t_{1L}}}, \tag{S19}\label{S19}
\end{align}
\begin{align}
\mathcal{T}(m)=\frac{t_{2}\sin[(m-1)\theta_\ell]}{\sin(\theta_\ell)}+\frac{\sqrt{t_{1L}t_{1R}}\sin(m\theta_\ell)}{\sin(\theta_\ell)}. \tag{S20}\label{S20}
\end{align}
The $\theta_\ell$ in Eqs. (\ref{S17}) and (\ref{S19}) is one of $M$ complex roots of the polynomial equation $\mathcal{T}(M+1)=0$ about $\cos(\theta_\ell)$, viz.,
\begin{align}
\mathcal{T}(M+1)=\frac{t_{2}\sin(M\theta_\ell)}{\sin(\theta_\ell)}
+\frac{\sqrt{t_{1L}t_{1R}}\sin[(M+1)\theta_\ell]}{\sin(\theta_\ell)}
=t_2U_{M-1}[\cos(\theta_\ell)]+\sqrt{t_{1L}t_{1R}}U_M[\cos(\theta_\ell)]=0, \tag{S21}\label{S21}
\end{align}
where $U_m(x)$ is a Chebyshev polynomial of the second kind, defined by the recursion formula $U_{m+1}(x)=2xU_m(x)-U_{m-1}(x)$ with $U_{0}(x)=1$ and $U_1(x) = 2x$. As an example, for $M=4$, $t_{1L}=t_2=2$ and $t_{1R}=1/2$, Eq. (\ref{S21}) reduces to $16\cos^4(\theta_\ell)+16\cos^3(\theta_\ell)-12\cos^2(\theta_\ell)-8\cos(\theta_\ell)+1=0$, resulting in four complex roots $\theta_{1,2,3,4}\simeq2.23$, $1.46$, $0.73$, and $\pi-0.69i$.

In a similar fashion, under $y$OBC, the exact solutions of  Eqs. (\ref{S8})--(\ref{S10}) can be found as
\begin{align}
\psi_{n,mA}=w^n\sin(n\phi_j)\varpi_\ell^{m},~~~~\psi_{n,mB}=(E-2Q)w^n\sin(n\phi_j)\varpi_\ell^{m},   \tag{S22}\label{S22}
\end{align}
\begin{align}
E=(\mu+1) Q \pm\sqrt{(\mu-1)^2 Q^2+T}\equiv E^{y\mathrm{OBC}}, \tag{S23}\label{S23}
\end{align}
where
\begin{equation}
Q=\sqrt{g_{1D}g_{1U}}\cos{\phi_j},~~~\phi_j=\frac{j\pi}{N+1},~~~w=\sqrt{\frac{g_{1U}}{g_{1D}}}. \tag{S24}\label{S24}
\end{equation}

As one can see, these solutions are very elegant in form: starting from the eigenenergy (\ref{S14}), which is easy to derive from the Bloch Hamiltonian, one can obtain the eigenenergy (\ref{S18}) for $E^{x\mathrm{OBC}}$ by making substitution $T\rightarrow R$ [or $\varpi_\ell\rightarrow r\exp(i\theta_\ell)$], and obtain the eigenenergy (\ref{S23}) for  $E^{y\mathrm{OBC}}$ by replacing $P\rightarrow 2Q$ only [or $\omega_j\rightarrow w\exp(i\phi_j)$]. The above solutions suggest that, under unidirectional OBC, there will definitely occur non-Hermitian skin effects, provided that either $w\neq1$ or $r\neq1$ is met.

\subsection{Exact closed-form solutions under double OBC}\label{Sec1C}

Lastly, under double OBC (dOBC), namely, $\kappa_{1,2}=\delta_{1,2}=0$, implying that the 2D lattice is fully open along both the $x$ and  $y$ directions, one can obtain the exact closed-form solutions of Eqs. (\ref{S8})--(\ref{S10}) as follows
\begin{align}
\psi_{n,mA}=w^n\sin(n\phi_j)r^{m-1}\mathcal{T}(m)\sin(\theta_\ell),~~~~\psi_{n,mB}=(E-2Q)w^n\sin(n\phi_j)r^m\sin(m\theta_\ell),   \tag{S25}\label{S25}
\end{align}
\begin{align}
E=(\mu+1)Q\pm\sqrt{(\mu-1)^2 Q^2+R}\equiv E^{\mathrm{dOBC}}. \tag{S26}\label{S26}
\end{align}
It is interesting to note that the eigenenergy (\ref{S26}) can be obtained from Eq. (\ref{S14}) by simultaneously making substitutions $P\rightarrow 2Q$ and $T\rightarrow R$ therein. Undoubtedly, under dOBC, when the parameter condition $|t_2/\sqrt{t_{1L}t_{1R}}|>1$ is satisfied \cite{Hou23}, the topological edge states pop up, with edge energies $E^{\mathrm{dOBC}}_\mathrm{edge}=2\mu Q$ or $2Q$. These edge energies can be obtained from Eq. (\ref{S26}) by letting $R\rightarrow 0$ therein. If $r\neq1$ and $w\neq1$ are further met, non-Hermitian skin effects take place along both $x$ and $y$ directions, resulting the formation of bulk corner states and skin-topological edge states, as shown in Fig. 1 in the main text.

\begin{figure}[h!]
\begin{center}
\includegraphics[width=16cm]{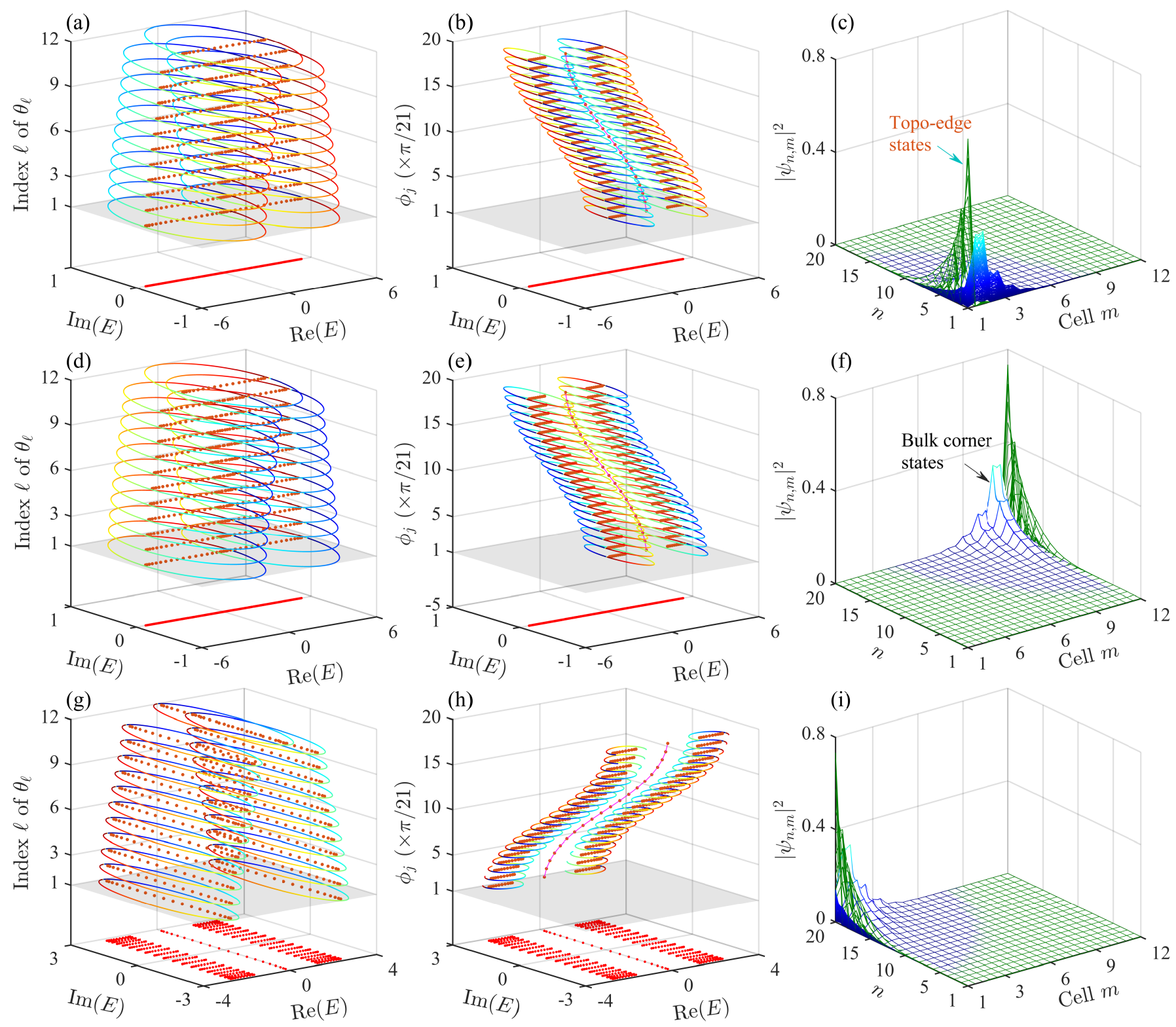}
\caption{\justify The bulk corner states and skin-topological edge states allowed in a bidirectionally open uniform ($\mu=1$) non-Hermitian SSH-HN lattice of size $M=12$ and $N=20$, for (a,b,c) $t_{1L}= 5/4$, $t_{1R}=1/4$, $g_{1D}=2$, $g_{1U}=1$, which implies $r<1$ and $w<1$; (d,e,f) $t_{1L}= 1/4$, $t_{1R}=5/4$, $g_{1D}=1$, $g_{1U}=2$, corresponding to $r>1$ and $w>1$; and (g,h,i) $t_{1L}= 5/4$, $t_{1R}=1/4$, $g_{1D}=i$, $g_{1U}=2i$, giving $r<1$ and $w>1$. The intercell hopping parameter $t_2=2$ is kept the same in all cases. } \label{figS1}
\end{center}
\end{figure}

Here we demonstrate in Supplementary Fig. \ref{figS1} that, for a bidirectionally open uniform 2D SSH-HN lattice where $\mu=1$, the skin-topological edge states (green lines) will occupy one side only, and tend to accumulate on one end of the side because of non-Hermitian skin effect. Besides, we show that the location of bulk corner states can be controlled by engineering the system parameters, which lies on the lower left corner for $w<1,~r<1$ [see Supplementary Figs. \ref{figS1}(a)--\ref{figS1}(c)], on the upper right corner for $w>1,~r>1$ [see Supplementary Figs. \ref{figS1}(d)--\ref{figS1}(f)], on the upper left corner for $w>1,~r<1$ [see Supplementary Figs. \ref{figS1}(g)--\ref{figS1}(i)], and on the lower right corner for $w<1,~r>1$ [see Fig. 2(e) in the main text].

Before proceeding further, we need to point out that our analytical solutions given in Secs. \ref{Sec1A}--\ref{Sec1C} are all in perfect agreement with numerical ones. As an example, we use the same parameters as in Figs. 2(e) and 2(f) in the main text and demonstrate in Supplementary Fig. \ref{figS2} the comparison between the analytical eigenstate solutions (\ref{S25}) and their numerical results (red solid circles). It is clearly seen that both solutions almost coincide with each other, as expected.

\begin{figure}[h!]
\begin{center}
\includegraphics[width=18cm]{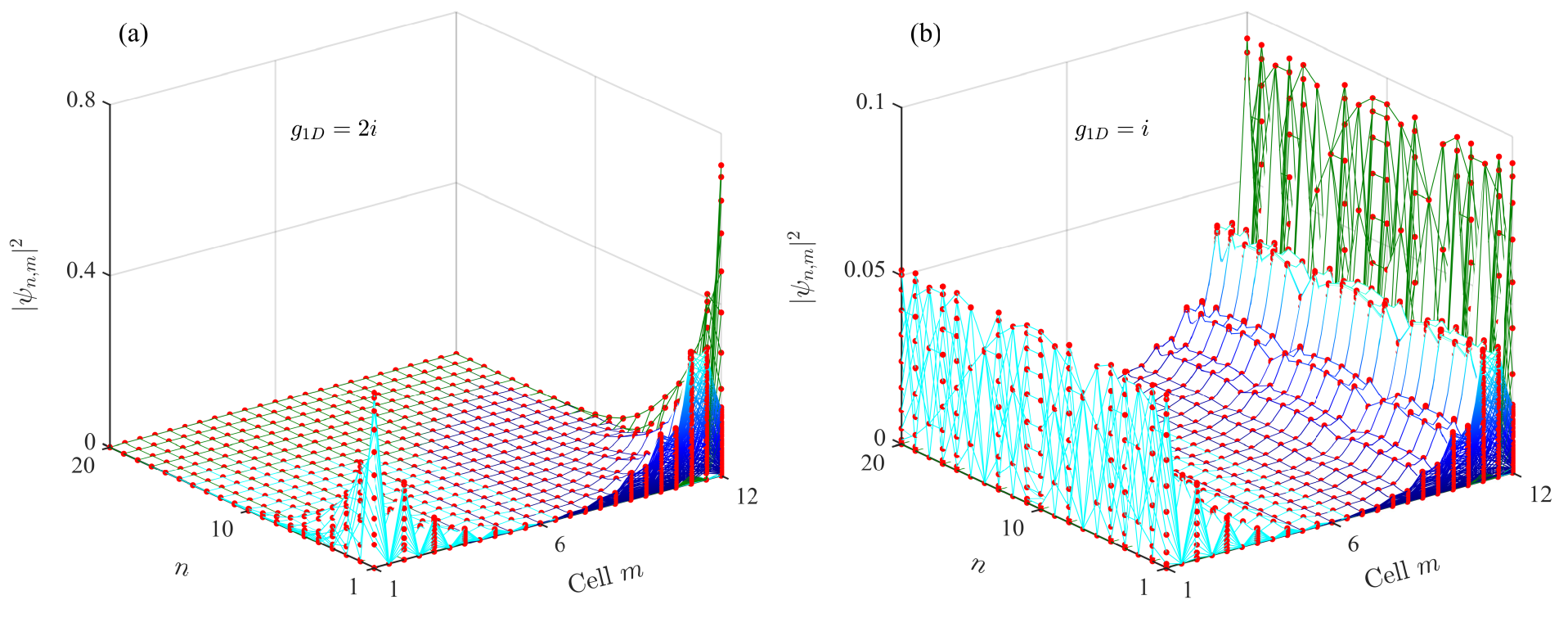}
\caption{\justify  Comparison between the analytical eigenstate solutions (\ref{S25}) and their numerical solutions (red solid circles) calculated from Hamiltonian (\ref{S3}) using the {\it Matlab} built-in command $\mathrm{eig}(H^{\mathrm{SN}})$, for two specific cases that have been considered in Fig. 2 in the main text: (a) the skin-topological edge states and the bulk corner states, and (b) the skin-effect-free topological edge states and the bulk skin modes. All parameter sets remain the same as in Figs. 2(e) and 2(f) in the main text. } \label{figS2}
\end{center}
\end{figure}

\section{The 2D SSH-SSH lattice model} \label{Sec2}

In the current section, let us provide the details of derivation of exact solutions for the 2D SSH-SSH lattice model illustrated in Fig. 1(b) in the main text \cite{Lee19,LiuF17,LiuT19}, under three different types of boundary conditions, namely, dGBC, unidirectional OBC, and dOBC.
The real-space Hamiltonian of this non-Hermitian rectangular lattice reads
\begin{align}
\hat{H}^{\mathrm{SS}}=&\sum_{n,m}(\hat{C}^{\dagger}_{n,m}M_{1}\hat{C}_{n,m}+\hat{C}^{\dagger}_{n,m+1}M_{2}^{\dagger}\hat{C}_{n,m}
+\hat{C}^{\dagger}_{n,m}M_{2}\hat{C}_{n,m+1}+\hat{C}^{\dagger}_{n+1,m}M_{3}^{\dagger}\hat{C}_{n,m}
+\hat{C}^{\dagger}_{n,m}M_{3}\hat{C}_{n+1,m}) \nonumber \\
&+\sum_{n}(\delta_1\hat{C}^{\dagger}_{n,1}M_{2}^{\dagger}\hat{C}_{n,M}
+\delta_2\hat{C}^{\dagger}_{n,M}M_{2}\hat{C}_{n,1})
+\sum_{m}(\kappa_1\hat{C}^{\dagger}_{1,m}M_{3}^{\dagger}\hat{C}_{N,m}
+\kappa_2\hat{C}^{\dagger}_{N,m}M_{3}\hat{C}_{1,m}),\tag{S27}\label{S27}
\end{align}
where
\begin{align}
	M_{1}=	\left[\begin{array}{cccc}
		0 &  t_{1L} &  g_{1D} &  0 \\
		t_{1R} &0   & 0  & g_{3D}    \\
        g_{1U} &0   & 0 &   t_{3L}   \\
        0 & g_{3U}  &   t_{3R}   & 0 \\		
	\end{array}\right],~~~
M_{2}=	\left[\begin{array}{cccc}
		0 &  0 &    0  &  0 \\
		t_{2} &0  & 0  &   0     \\
0&0   & 0 &   0   \\
0   &  0  &   t_{4}   & 0 \\		
	\end{array}\right],~~~
M_{3}=	\left[\begin{array}{cccc}
		0 &  0 &    0  &  0 \\
		0 &0  & 0&    0    \\
g_{2}  &0   & 0 &   0   \\
0 & g_{4}   &   0   & 0 \\		
	\end{array}\right], \tag{S28}\label{S28}
\end{align}
and $\hat{C}^{\dagger}_{n,m} = (\hat{a}^{\dagger}_{n,m}, \hat{b}^{\dagger}_{n,m}, \hat{c}^{\dagger}_{n,m},\hat{d}^{\dagger}_{n,m})$ are the creation operators of particles on sublattices A, B, C, and D of the cell at the spatial coordinate $(n,m)$, that is, in the $n$th row and $m$th column of the lattice.

Analogous to the SSH-HN lattice situation, we can rewrite the 2D Hamiltonian (\ref{S27}) as a $4MN\times4MN$ matrix:
\begin{align}
	H^{\mathrm{SS}}=	\left[\begin{array}{ccccccc}
K_{1}& K_{D}&            &            &              &            & \kappa_1 K_{3}       \\
K_{U}& K_{2}& K_{3}&            &              &            &  \\
          & K_{3} &K_{1}& K_{D} &             &           & \\
          &             & K_{U}& K_{2} & K_{3} &           &     \\
          &              &            & K_{3} & K_{1} & \ddots   &   \\
          &	           &            &             &  \ddots       & \ddots &K_{D}  \\
\kappa_2 K_{3}&   &            &             &              &K_{U}  &K_{2} 	
	\end{array}\right]_{4MN\times4MN},   \tag{S29}\label{S29}
\end{align}
where $\kappa_1 K_{3}$ and $\kappa_2 K_{3}$ at two corners represent the boundary conditions along the $y$ direction, and $K_{1,2,3}$ and $K_{X}$ $(X=U,D)$ are $2M\times2M$  matrices composed of hopping parameters, defined by
\begin{align}
	K_{1}=	\left[\begin{array}{cccccc}
0& t_{1L}& & & & \delta_1 t_{2}  \\
t_{1R} &0          &t_{2} &  & &    \\
                 &t_{2}   &0        &t_{1L} & &  \\
                 &                   &t_{1R} &0  &\ddots & \\
	&	&                        &   \ddots     &\ddots &t_{1L}        \\
	\delta_2 t_{2}  &        &   &      &t_{1R}  &0 \\		
	\end{array}\right]_{2M\times2M},~~~
	K_{2}=	\left[\begin{array}{cccccc}
0& t_{3L}& & &  &  \delta_1 t_{4}  \\
t_{3R} &0          &t_{4} &   & &    \\
                 &t_{4}   &0        &t_{3L} &  &  \\
                 &                   &t_{3R} &0  &\ddots  & \\
	&	&                        &   \ddots     &\ddots &t_{3L}        \\
	\delta_2 t_{4}  &       &   &      &t_{3R}  &0 \\		
	\end{array}\right]_{2M\times2M},  \tag{S30}\label{S30}
\end{align}
\begin{align}
	K_{3}=	\left[\begin{array}{cccccc}
		g_{2}&   &   &          &     &        \\
		 &g_{4}  &  &     &    &          \\
		          &   & g_{2} &   &  &    \\
		          &   & & g_{4} &   &      \\
		          &          &    &        &\ddots &     \\
		  &           &  &    & &g_{4}  \\		
	\end{array}\right]_{2M\times2M},~~~~
	K_{X}=	\left[\begin{array}{cccccc}
		g_{1X} &   &             &     &  &       \\
		 &g_{3X}  &  &     &    &          \\
		          &   & g_{1X} &   &   &   \\
		          &   & & g_{3X} &   &      \\
		          &        &   &          &\ddots &     \\
		  &           &  &    & &g_{3X}  \\		
	\end{array}\right]_{2M\times2M}.  \tag{S31}\label{S31}
\end{align}
Now the eigenvector $|\Psi^{\mathrm{SS}}\rangle$ of the Hamiltonian (\ref{S27}) can be arranged into the following column array:
\begin{align}
|\Psi^{\mathrm{SS}}\rangle = \left[\begin{array}{c}
|\psi_{1}\rangle \\
|\psi_{2}\rangle\\
|\psi_{3}\rangle \\
\vdots \\
|\psi_{N}\rangle
\end{array}\right], \tag{S32}\label{S32}
\end{align}
where
\begin{align}
|\psi_{n}\rangle =
[\underbrace{\psi_{n,1A},\psi_{n,1B},\cdots,\psi_{n,mA},\psi_{n,mB},\cdots,\psi_{n,MA},\psi_{n,MB},
\psi_{n,1C},\psi_{n,1D},\cdots,\psi_{n,mC},\psi_{n,mD},\cdots,\psi_{n,MC},\psi_{n,MD}}_{4M}]^{\mathrm{T}},\tag{S33}\label{S33}
\end{align}
gives the state components in the $n$th row of the rectangular lattice.

Then, substituting Eqs. (\ref{S29}) and (\ref{S32}) into $H^{\mathrm{SS}}|\Psi^{\mathrm{SS}}\rangle=E|\Psi^{\mathrm{SS}}\rangle$ followed by algebraic manipulations, one can obtain a system of bulk equations
\begin{align}
\left[\begin{array}{cc}
		K_1 &  K_{D}  \\
		K_{U} &K_2    \\		
	\end{array}\right]|\psi_{n}\rangle+\left[\begin{array}{cc}
		0 &  0  \\
		K_{3} &0    \\		
	\end{array}\right]|\psi_{n+1}\rangle+\left[\begin{array}{cc}
		0 &  K_{3}  \\
		0 &0    \\		
	\end{array}\right]|\psi_{n-1}\rangle=E|\psi_{n}\rangle,  \tag{S34}\label{S34}
\end{align}
where $n=2,\cdots,N-1$, and two boundary equations given by
\begin{align}
\left[\begin{array}{cc}
		K_1 &  K_{D}  \\
		K_{U} &K_2    \\		
	\end{array}\right]|\psi_{1}\rangle+\left[\begin{array}{cc}
		0 &  0  \\
		K_{3} &0    \\		
	\end{array}\right]|\psi_{2}\rangle+ \left[\begin{array}{cc}
		0 &  \kappa_1K_{3}  \\
		0 &0    \\		
	\end{array}\right]|\psi_{N}\rangle=&E|\psi_{1}\rangle, \tag{S35}\label{S35}
\end{align}
\begin{align}
\left[\begin{array}{cc}
		K_1 &  K_{D}  \\
		K_{U} &K_2    \\		
	\end{array}\right]|\psi_{N}\rangle+ \left[\begin{array}{cc}
		0 &  0  \\
		\kappa_2K_{3} &0    \\		
	\end{array}\right]|\psi_{1}\rangle+\left[\begin{array}{cc}
		0 &  K_{3}  \\
		0 &0    \\		
	\end{array}\right]|\psi_{N-1}\rangle=&E|\psi_{N}\rangle.  \tag{S36}\label{S36}
\end{align}
Here, Eqs. (\ref{S34})--(\ref{S36}) have been expressed as a compact block matrix form, for the sake of conciseness. Besides, we assume that $t_{3Z}/t_{1Z}=t_4/t_2=\mu$ and $g_{3X}/g_{1X}=g_4/g_2=\nu$, where $Z=L,R$ and $X=U,D$. Under the circumstances, Eqs. (\ref{S34})--(\ref{S36}) can be exactly solved, no matter what boundary conditions are imposed.

Likewise, implementing the Fourier transformation of the Hamiltonian (\ref{S27}), one can also obtain the Bloch Hamiltonian in reciprocal momentum space \cite{Shen13}:
\begin{align}
\mathcal{H}^{\mathrm{SS}}(k_{x},k_{y})=	\left[\begin{array}{cccc}
0  & t_{1L}+t_{2}\exp(-ik_{x}) &g_{1D}+g_{2}\exp(-ik_{y})& 0 \\
t_{1R}+t_{2}\exp(ik_{x})  &  0 & 0 &  g_{3D}+g_{4}\exp(-ik_{y}) \\
g_{1U}+g_{2}\exp(ik_{y})& 0 &0  &t_{3L}+t_{4}\exp(-ik_{x})\\
 0  & g_{3U}+g_{4}\exp(ik_{y}) & t_{3R}+t_{4}\exp(ik_{x}) &0
\end{array}\right].   \tag{S37}\label{S37}
\end{align}
From this Bloch Hamiltonian, one can easily prove that the above 2D SSH-SSH lattice will respect the sublattice symmetry (SLS). As a result, its eigenenergies will come in pairs ($E,~-E$) \cite{Kawa19,Li24}.

\subsection{Exact solutions under dGBC }\label{Sec2A}

Let us once again start with the simplest dGBC, which corresponds to $\delta_1=\delta_2^{-1}=e^{i\varphi}$ and $\kappa_1=\kappa_2^{-1}=e^{i\vartheta}$. Under this boundary condition,  Eqs. (\ref{S34})--(\ref{S36}) are found to admit the following exact solutions:
\begin{align}
\psi_{n,mA}=\omega_j^{n-1}\varpi_\ell^{m},~~~\psi_{n,mB}=E_{B}\omega_j^{n-1}\varpi_\ell^{m}, \tag{S38}\label{S38}
\end{align}
\begin{align}
\psi_{n,mC}=E_{C}\omega_j^{n-1}\varpi_\ell^{m},~~~\psi_{n,mD}=E_{D}\omega_j^{n-1}\varpi_\ell^{m}, \tag{S39}\label{S39}
\end{align}
\begin{align}
E=\pm\sqrt{\frac{(\mu^2+1)T+(\nu^2+1)G\pm\sqrt{\Delta(T,G)}}{2}}\equiv E^{\mathrm{dGBC}},   \tag{S40}\label{S40}
\end{align}
where $\varpi_\ell=\exp\left(\frac{i2\ell\pi}{M}-\frac{i\varphi}{M}\right)$, $\omega_j=\exp\left(\frac{i2j\pi}{N}-\frac{i\vartheta}{N}\right)$, exactly the same as given by Eq. (\ref{S16}), and
\begin{align}
T=(t_{1R}+t_{2}\varpi_\ell)\left(t_{1L}+\frac{t_{2}}{\varpi_\ell}\right),~~~
G=(g_{1U}+g_{2}\omega_j)\left(g_{1D}+\frac{g_{2}}{\omega_j}\right), \tag{S41}\label{S41}
\end{align}
\begin{align}
\Delta(T,G)=\left[(\mu+1)^2T+(\nu-1)^2G\right]\left[(\mu-1)^2T+(\nu+1)^2G\right]. \tag{S42}\label{S42}
\end{align}
The other coefficients $E_B$, $E_{C}$, and $E_{D}$ in Eqs. (\ref{S38}) and (\ref{S39}) are defined by
\begin{align}
E_{B}=\frac{
\nu E^2-\nu G+\mu T}{(\mu+\nu)(t_{1L}+t_{2}/\varpi_\ell)E},~~~~
E_{C}=\frac{
\mu E^2+\nu G-\mu T}{(\mu+\nu)(g_{1D}+g_{2}/\omega_j)E},~~~~
E_{D}=\frac{E^2-T-G}{(\mu+\nu)(t_{1L}+t_{2}/\varpi_\ell)(g_{1D}+g_{2}/\omega_j)}.\tag{S43}\label{S43}
\end{align}
It should be noted that the subscripts $j=1,\cdots,N$ and $\ell=1,\cdots,M$ in Eqs. (\ref{S38}) and (\ref{S39}) are used for labelling the eigenstates, which will have a total number of $4M\times N$. Besides, the solutions for dPBC follow easily by setting $\varphi=\vartheta=0$.

\subsection{Exact closed-form solutions under unidirectional OBC}\label{Sec2B}

As done in Sec. \ref{Sec1B}, we subdivide the unidirectional OBC into $x$OBC and $y$OBC. Under $x$OBC, which corresponds to $\kappa_1=\kappa_2^{-1}=e^{i\vartheta}$ and $\delta_1=\delta_2=0$ (i.e., OBC in $x$ but PBC in $y$ dimension), one can obtain the exact eigenstate solution:
\begin{align}
\psi_{n,mA}=\omega_j^{n-1}r^{m-1}\mathcal{T}(m)\sin(\theta_\ell),~~~\psi_{n,mB}=E_{B}\omega_j^{n-1}r^m\sin(m\theta_\ell), \tag{S44}\label{S44}
\end{align}
\begin{align}
\psi_{n,mC}=E_{C}\omega_j^{n-1}r^{m-1}\mathcal{T}(m)\sin(\theta_\ell),~~~\psi_{n,mD}=E_{D}\omega_j^{n-1}r^m\sin(m\theta_\ell), \tag{S45}\label{S45}
\end{align}
associated with the eigenenergy:
\begin{align}
E=\pm\sqrt{\frac{(\mu^2+1)R+(\nu^2+1)G\pm\sqrt{\Delta(R,G)}}{2}}\equiv E^{x\mathrm{OBC}},   \tag{S46}\label{S46}
\end{align}
where
\begin{align}
R=2t_{2}\sqrt{t_{1L}t_{1R}}\cos(\theta_\ell)+t_{1L}t_{1R}+t_{2}^2, ~~~~r=\sqrt{\frac{t_{1R}}{t_{1L}}}, \tag{S47}\label{S47}
\end{align}
\begin{align}
\mathcal{T}(m)=\frac{t_{2}\sin[(m-1)\theta_\ell]}{\sin(\theta_\ell)}+\frac{\sqrt{t_{1L}t_{1R}}\sin(m\theta_\ell)}{\sin(\theta_\ell)}, \tag{S48}\label{S48}
\end{align}
\begin{align}
\Delta(R,G)=\left[(\mu+1)^2R+(\nu-1)^2G\right]\left[(\mu-1)^2R+(\nu+1)^2G\right]. \tag{S49}\label{S49}
\end{align}
The $\theta_\ell$ in Eqs. (\ref{S44})--(\ref{S48}) is one of the $M$ complex roots of the polynomial equation $\mathcal{T}(M+1)=0$. We note that Eqs. (\ref{S47}) and (\ref{S48}) are identical to Eqs. (\ref{S19}) and (\ref{S20}) used in the SSH-HN lattice model, as the same set of hopping parameters is adopted.
Now the coefficients $E_B$, $E_{C}$, and $E_{D}$ in Eqs. (\ref{S44}) and (\ref{S45}) become
\begin{align}
E_{B}=\frac{
\nu E^2-\nu G+\mu R}{(\mu+\nu)E},~~~~
E_{C}=\frac{
\mu E^2+\nu G-\mu R}{(\mu+\nu)(g_{1D}+g_{2}/\omega_j)E},~~~~
E_{D}=\frac{E^2-G-R}{(\mu+\nu)(g_{1D}+g_{2}/\omega_j)}.\tag{S50}\label{S50}
\end{align}

Similarly, under $y$OBC (that is, OBC in $y$ but PBC in $x$ dimension), one can get the eigenenergy as follows
\begin{align}
E=\pm\sqrt{\frac{(\mu^2+1)T+(\nu^2+1)S\pm\sqrt{\Delta(T,S)}}{2}}\equiv E^{y\mathrm{OBC}},   \tag{S51}\label{S51}
\end{align}
where
\begin{align}
S=2 g_{2} \sqrt{g_{1 D} g_{1 U}}\cos(\phi_j)+g_{1 D} g_{1 U}+g_{2}^{2}, \tag{S52}\label{S52}
\end{align}
\begin{align}
\Delta(T,S)=\left[(\mu+1)^2T+(\nu-1)^2S\right]\left[(\mu-1)^2T+(\nu+1)^2S\right]. \tag{S53}\label{S53}
\end{align}
The $\phi_j$ in Eq. (\ref{S52}) is one of the $N$ complex roots of the polynomial equation $\mathcal{G}(N+1)=0$, where
\begin{align}
\mathcal{G}(n)=\frac{g_{2}\sin[(n-1)\phi_j]}{\sin(\phi_j)}+\frac{\sqrt{g_{1 D} g_{1 U}}\sin(n\phi_j)}{\sin(\phi_j)}. \tag{S54}\label{S54}
\end{align}
The eigenvector corresponding to $E^{y\mathrm{OBC}}$ can also be derived easily, with its state components given by
\begin{align}
\psi_{n,mA}=w^{n-1}\mathcal{G}(n)\sin(\phi_j)\varpi^m,
~~~\psi_{n,mB}=E_{B}w^{n-1}\mathcal{G}(n)\sin(\phi_j)\varpi^m, \tag{S55}\label{S55}
\end{align}
\begin{align}
\psi_{n,mC}=E_{C}w^n\sin(n\phi_j)\varpi^m,
~~~\psi_{n,mD}=E_{D}w^n\sin(n\phi_j)\varpi^m, \tag{S56}\label{S56}
\end{align}
where $w=\sqrt{g_{1U}/g_{1D}}$ and
\begin{align}
E_{B}=\frac{
\nu E^2-\nu S+\mu T}{(\mu+\nu)(t_{1L}+t_{2}/\varpi_\ell)E},~~~~
E_{C}=\frac{
\mu E^2+\nu S-\mu T}{(\mu+\nu)E},~~~~
E_{D}=\frac{E^2-T-S}{(\mu+\nu)(t_{1L}+t_{2}/\varpi_\ell)}.\tag{S57}\label{S57}
\end{align}

\begin{figure}[h!]
\begin{center}
\includegraphics[width=16cm]{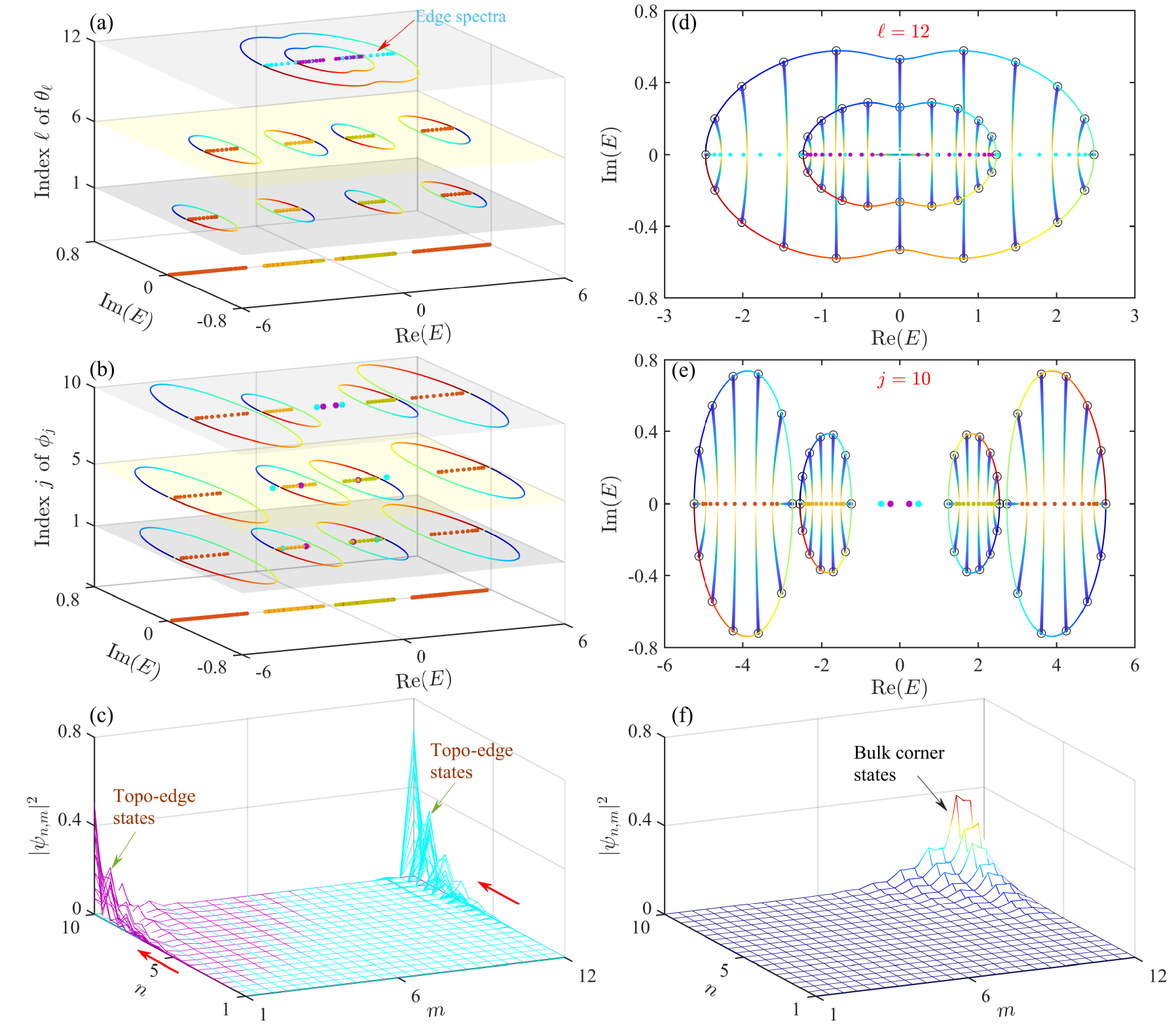}
\caption{\justify The bulk corner states and skin-topological edge states formed in a fully open 2D nonuniform non-Hermitian SSH-SSH lattice, for $t_{1L}= 1/4$, $t_{1R}=1$, $t_2=2$, $g_{1D}=2/5$, $g_{1U}=6/5$, $g_2=1/2$, and $\mu=\nu=2$, which satisfies the line-gap topology condition $|t_2/\sqrt{t_{1R}t_{1L}}|>1$ for the horizontal chain. (a,b) Energy spectra; (d,e) enlarged views of top planes in (a,b); and (c,f) the gapped topological edge states and the bulk corner states, all have been normalized by total intensity.} \label{figS3}
\end{center}
\end{figure}

\subsection{Exact closed-form solutions under dOBC }\label{Sec2C}

Finally, under dOBC, where $\kappa_{1,2}=\delta_{1,2}=0$ (i.e., keeping OBC in both $x$ and $y$ dimensions), one obtains the exact eigenstate solution of Eqs. (\ref{S34})--(\ref{S36}) as
\begin{align}
\psi_{n,mA}=w^{n-1}\mathcal{G}(n)\sin(\phi_j)r^{m-1}\mathcal{T}(m)\sin(\theta_\ell),
~~~\psi_{n,mB}=E_{B}w^{n-1}\mathcal{G}(n)\sin(\phi_j)r^m\sin(m\theta_\ell), \tag{S58}\label{S58}
\end{align}
\begin{align}
\psi_{n,mC}=E_{C}w^n\sin(n\phi_j)r^{m-1}\mathcal{T}(m)\sin(\theta_\ell),
~~~\psi_{n,mD}=E_{D}w^n\sin(n\phi_j)r^m\sin(m\theta_\ell), \tag{S59}\label{S59}
\end{align}
associated to the eigenenergy
\begin{align}
E=\pm\sqrt{\frac{(\mu^2+1)R+(\nu^2+1)S\pm\sqrt{\Delta(R,S)}}{2}}\equiv E^{\mathrm{dOBC}},   \tag{S60}\label{S60}
\end{align}
where
\begin{align}
\Delta(R,S)=\left[(\mu+1)^2R+(\nu-1)^2S\right]\left[(\mu-1)^2R+(\nu+1)^2S\right]. \tag{S61}\label{S61}
\end{align}
The coefficients $E_B$, $E_{C}$, and $E_{D}$ in Eqs. (\ref{S58}) and (\ref{S59}) now take the forms
\begin{align}
E_{B}=\frac{(\mu\nu+1)R E}{E^2+\mu\nu R-\nu^2S},~~~~
E_{C}=\frac{(E^2-R-\nu^2S)E}{E^2+\mu\nu R-\nu^2S},~~~~
E_{D}=\frac{(\mu E^2-\mu R+\nu S)R}{E^2+\mu\nu R-\nu^2S}.\tag{S62}\label{S62}
\end{align}

As one can verify, for a nonuniform 2D lattice with $\mu\neq1$ and $\nu\neq1$, when $|t_2/\sqrt{t_{1L}t_{1R}}|>1$ is met, the topological edge states pop up on left and right sides, with eigenenergies $E^{\mathrm{dOBC}}_\mathrm{edge}=\pm\nu\sqrt{S}$ or $\pm\sqrt{S}$, obtained from Eq. (\ref{S60}) by setting $R\rightarrow0$ therein. If further the parameter condition $|g_2/\sqrt{g_{1 D} g_{1 U}}|>1$ is satisfied, the gapped edge states could emerge on four sides of 2D lattice, with eigenenergies $E^{\mathrm{dOBC}}_\mathrm{edge}=\pm\nu\sqrt{S}$, $\pm\sqrt{S}$, $\pm\mu\sqrt{R}$, or $\pm\sqrt{R}$, obtained from Eq. (\ref{S60}) by taking the limit $R\rightarrow0$ and $S\rightarrow0$, respectively. One can refer to the next Sec. \ref{Sec3} for detailed information about these topological edge states.  A close inspection of these analytical solutions reveals that, while the gapped edge states on the right side are given by the non-decaying wave solutions on sublattices B and D, those on the left side are dominated by the wave solutions on sublattices A and C, since now $E_B\simeq0$ and $E_D\simeq0$ in Eq. (\ref{S62}). Likewise, the topological edge states formed on front side and the ones on back side are attributed to the dominant state distributions on sublattices (A,B) and (C,D), respectively. Of course, when taking both $R\rightarrow0$ and $S\rightarrow0$ in Eq. (\ref{S60}), one obtains a zero energy, i.e., $E^{\mathrm{dOBC}}_\mathrm{corner}=0$, which implies that there are now emerging the topological corner states which will stand on the corners of the 2D lattice.

Here we demonstrate in Supplementary Fig. \ref{figS3} the bulk corner states and skin-topological edge states in a fully open uniform  non-Hermitian SSH-SSH lattice, when the system parameters fulfil $|t_2/\sqrt{t_{1L}t_{1R}}|>1$ but $|g_2/\sqrt{g_{1D}g_{1U}}|<1$. Obviously, in this situation, the topological edge states will gather on the left and right sides of 2D lattice, but experience non-Hermitian skin effect along the positive $y$ direction, as indicated by red arrows in Supplementary Fig. \ref{figS3}(c).

\begin{figure}[h!]
\begin{center}
\includegraphics[width=16cm]{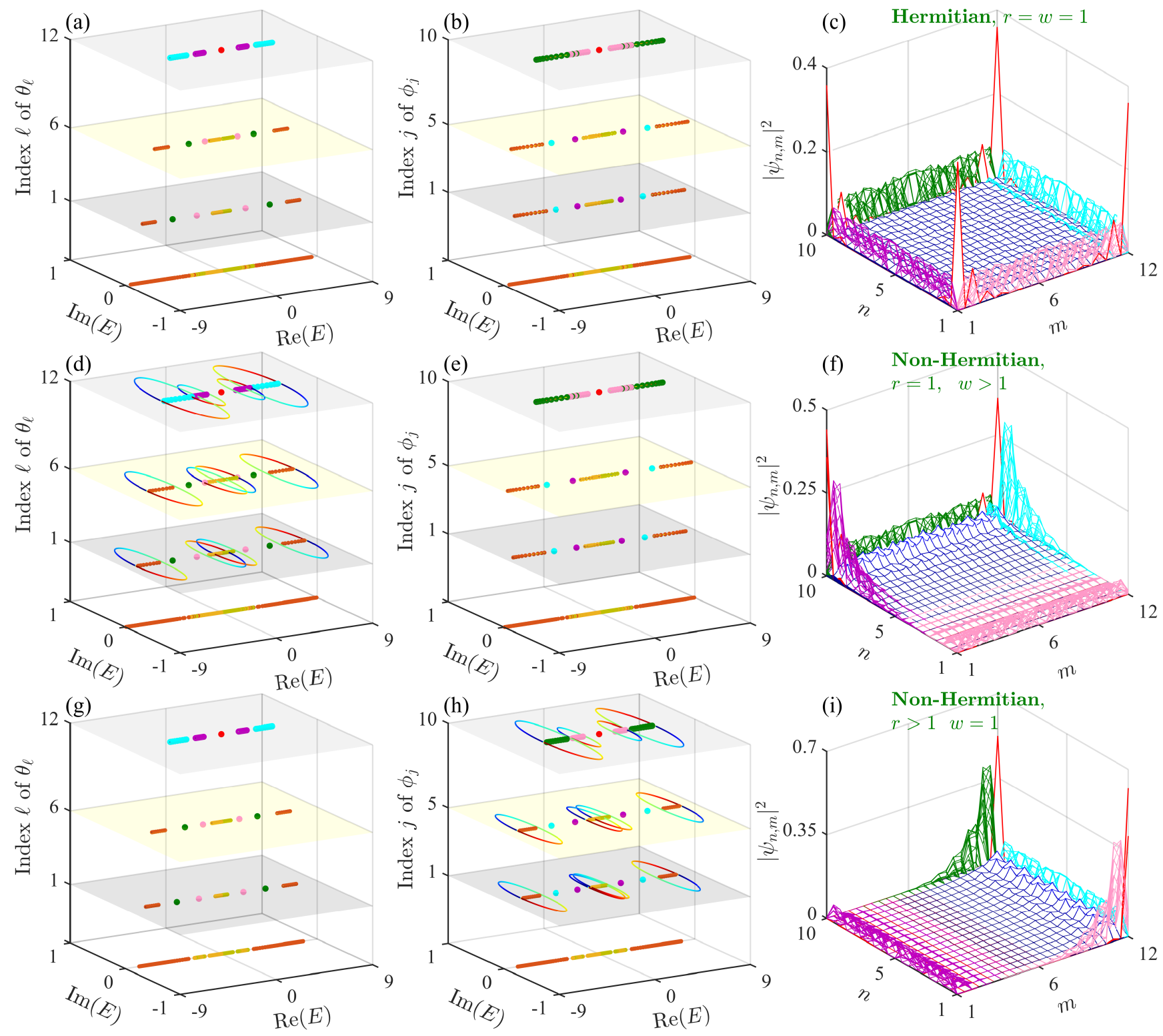}
\caption{\justify Interplay between topological edge states and non-Hermitian skin effect in a fully open 2D nonuniform SSH-SSH lattice, for three sets of system parameters. (a,b,c) The Hermitian case where $t_{1L}=t_{1R}= 1$, $g_{1D}=g_{1U}=2/5$, implying $r=w=1$; (d,e,f) non-Hermitian case, where $t_{1L}=t_{1R}= 1$, $g_{1D}=2/5$, $g_{1U}=6/5$, implying $w>1$ but $r=1$; and (g,h,i) non-Hermitian case, where $t_{1L}=1/4$, $t_{1R}= 1$, $g_{1D}=g_{1U}=2/5$, implying $r>1$ and $w=1$. The other parameters are kept the same, namely, $t_2=g_2=\mu=\nu=2$.
The Hermitian case has also been presented as Figs. 3(a,b,c) in the main text.} \label{figS4}
\end{center}
\end{figure}

In Supplementary Fig. \ref{figS4}, we demonstrate the interesting real-space dynamics allowed in a nonuniform SSH-SSH lattice with dOBC, using three sets of system parameters, which all satisfy $|t_2/\sqrt{t_{1L}t_{1R}}|>1$ and $|g_2/\sqrt{g_{1D}g_{1U}}|>1$ simultaneously. In this situation, the topological edge states pop up and build on four sides of the rectangular lattice.  As one can see, in the Hermitian case where $r=w=1$, there appear four zero-energy topological corner states locating on four corners, each surrounded by two gapped edge states on adjoining edges, each with intensity around $(1-t_{1L}t_{1R}/t_2^2)(1-g_{1D}g_{1U}/g_2^2)/2=0.36$ [see Supplementary Figs. \ref{figS4}(a,b,c) here or Figs. 3(a,b,c) in the main text]. However, when we set $w>1$ but $r=1$ (non-Hermitian),  the topological edge states on the left and right sides undergo non-Hermitian skin effect but those on the front and back sides remain distributed extensively, leading to the disappearance of topological corner states on the front two corners [see Supplementary Figs. \ref{figS4}(d,e,f)]. By the same token, if we make $r>1$ but $w=1$ (non-Hermitian), the topological corner states would disappear from the left two corners [see Supplementary Figs. \ref{figS4}(g,h,i)]. In the latter two non-Hermitian cases, the topological corner states possess a stronger intensity equal to $0.44$ or $0.62$. Particularly, if we take $r>1$ and $w>1$, only one single degenerate topological corner state remains, which has an intensity around 0.94, much larger than that in Hermitian case [see Fig. 3(f) in the main text]. This intensity is also inaccessible to the usual bulk corner states which have a maximum intensity around $0.29$ for the same set of system parameters used [see Fig. 3(f) in the main text].

As a concluding remark of this section, we emphasize once again that there is almost no inconsistency between our analytical solutions obtained above and the numerical results calculated from the {\it Matlab} built-in command $\mathrm{eig}(H^{\mathrm{SS}})$, as indicated in Supplementary Fig. \ref{figS5}, where we used the same parameters as in Figs. 3(c) and 3(f) in the main text. Obviously, compared with the numerical solutions which always treat the eigenstates as a whole, our analytical solutions enable one to distinguish the bulk states, the skin-topological edge states, and the in-edge corner states from each other. 

\begin{figure}[h!]
\begin{center}
\includegraphics[width=16cm]{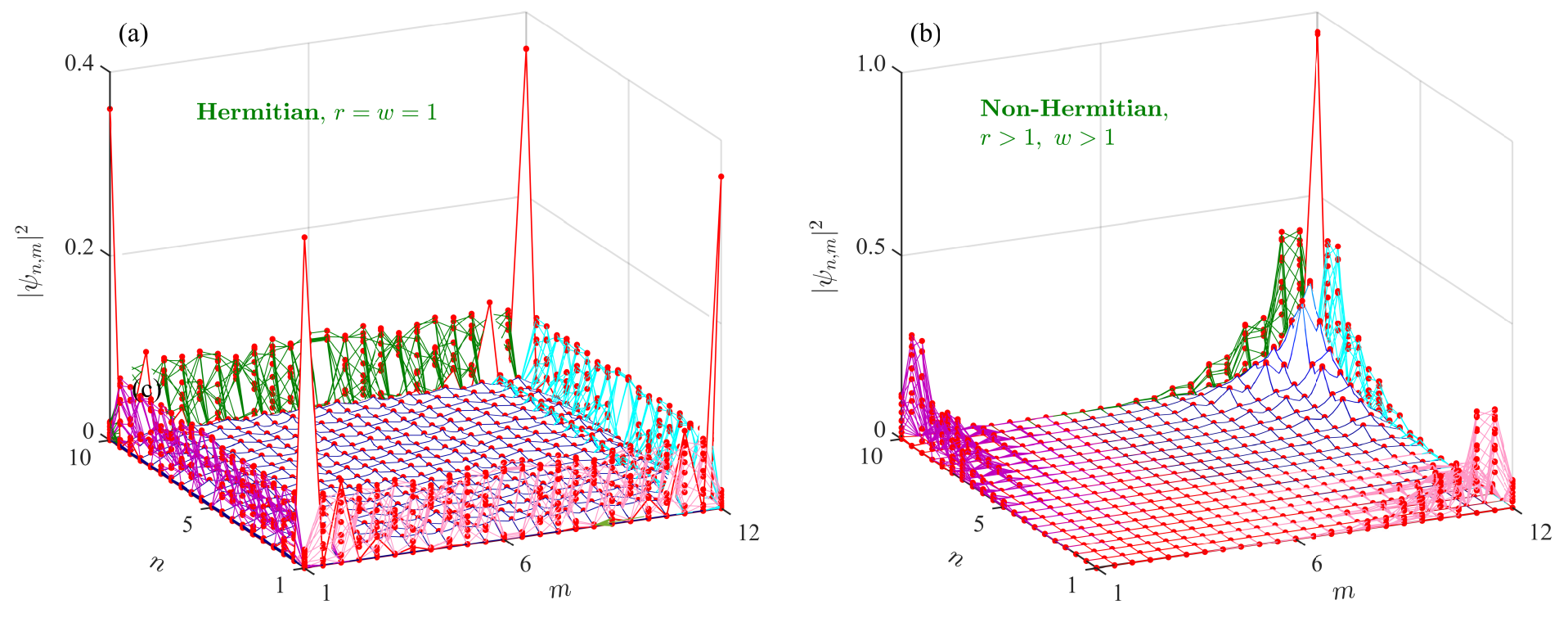}
\caption{Comparison between analytical solutions (\ref{S58}) and (\ref{S59}) and numerical results (red solid circles) for (a) the Hermitian case, (b) the non-Hermitian case, using the same parameters as in Figs. 3(c) and 3(f) in the main text.  } \label{figS5}
\end{center}
\end{figure}

\section{Generalized Brillouin zones, topological edge-state conditions, and phase diagrams} \label{Sec3}

In this section, we first obtain the explicit formulas defining the generalized Brillouin zone (GBZ) of the bulk states in a 2D lattice under dOBC, and then derive the parameter conditions for topological edge states that can occur in such fully open 2D lattice, starting from the general solutions (\ref{S58})--(\ref{S60}). Finally, we provide the phase diagrams to show the hopping dynamics of these topological edge states. For brevity, here we only take the complicated SSH-SSH lattice as an example. The results obtained here can also be applied to the 2D SSH-HN model.

\subsection{Generalized Brillouin zone} \label{Sec3A}

To obtain the GBZ formulas, let us rewrite the momentum-space Hamiltonian (\ref{S37}) for 2D SSH-SSH lattice as
\begin{align}
\mathcal{H}^{\mathrm{SS}}(\beta_{x},\beta_{y})=	\left[\begin{array}{cccc}
0  & t_{1L}+t_{2}/\beta_{x} &g_{1D}+g_{2}/\beta_{y}& 0 \\
t_{1R}+t_{2}\beta_{x}  &  0 & 0 &  \nu(g_{1D}+g_{2}/\beta_{y}) \\
g_{1U}+g_{2}\beta_{y}& 0 &0  &\mu(t_{1L}+t_{2}/\beta_{x})\\
 0  & \nu(g_{1U}+g_{2}\beta_{y}) & \mu(t_{1R}+t_{2}\beta_{x}) &0
\end{array}\right],   \tag{S63}\label{S63}
\end{align}
where $\beta_{x}=\exp(ik_x)$ and $\beta_{y}=\exp(ik_y)$ ($k_{x,y}\in \mathds{C}$), and meanwhile we have inserted back $t_{3L}=\mu t_{1L}$, $t_{3R}=\mu t_{1R}$, $t_4=\mu t_2$, $g_{3D}=\nu g_{1D}$, $g_{3U}=\nu g_{1U}$, and $g_4=\nu g_2$. It is worth noting that, under dOBC, the trajectories of $\beta_{x}$ and $\beta_{y}$ in the complex plane are no longer equal to the Brillouin zone (BZ) traced by $\exp(ik_x)$ and $\exp(ik_y)$ ($k_{x,y}\in \mathds{R}$), respectively, as now $k_{x,y}$ become complex when dOBC is imposed. It is also impracticable to get the GBZs by solving numerically the characteristic equation det$[E-H^{\mathrm{SS}}(\beta_{x}, \beta_{y})] =0$, as the latter involves two free variables $\beta_x$ and $\beta_y$, although the energy $E$ is easily obtained numerically.

As a matter of fact, one can solve the characteristic equation det$[E-H^{\mathrm{SS}}(\beta_{x}, \beta_{y})] =0$ analytically, and get the explicit formulas of $\beta_x$ and $\beta_y$ using our energy solution (\ref{S60}). To be specific, we transform det$[E-H^{\mathrm{SS}}(\beta_{x}, \beta_{y})] =0$ equivalently into
\begin{align}
E^2(1+\wp)(1+\nu^2\wp)-\left(t_{1L}+\frac{t_2}{\beta_x}\right)(t_{1R}+t_2\beta_x)(1-\mu\nu\wp)^2=0, \tag{S64}\label{S64}
\end{align}
where
\begin{align}
\wp=\frac{(g_{1D}+g_2/\beta_y)(g_{1U}+g_2\beta_y)}{\mu^2(t_{1L}+t_2/\beta_x)(t_{1R}+t_2\beta_x )-E^2}. \tag{S65}\label{S65}
\end{align}
Then, substituting the exact solution (\ref{S60}) for $E$ into Eq. (\ref{S64}) followed by simplifications, one obtains exactly
\begin{align}
\beta_x=r\exp(\pm i\theta_\ell),~~~~\beta_y=w\exp(\pm i\phi_j), \tag{S66}\label{S66}
\end{align}
where $r=\sqrt{t_{1R}/t_{1L}}$, $w=\sqrt{g_{1U}/g_{1D}}$, and $\theta_\ell$ and $\phi_j$ are determined, respectively, by the following two polynomial equations
\begin{align}
\mathcal{T}(M+1)=t_2U_{M-1}[\cos(\theta_{\ell})]+\sqrt{t_{1L}t_{1R}}U_M[\cos(\theta_\ell)]=0, \tag{S67}\label{S67}
\end{align}
\begin{align}
\mathcal{G}(N+1)=g_2U_{N-1}[\cos(\phi_j)]+\sqrt{g_{1D}g_{1U}}U_N[\cos(\phi_j)]=0. \tag{S68}\label{S68}
\end{align}
Here $U_i(x)$ is a Chebyshev polynomial of the second kind satisfying the recursion formula $U_{i+1}(x)=2xU_i(x)-U_{i-1}(x)$, with $U_{0}(x)=1$ and $U_1(x) = 2x$, as stated in Sec. \ref{Sec1B}. Therefore, for given set of hopping parameters and arbitrary lattice size $2M\times 2N$, one can get all $\theta_\ell$ and $\phi_j$ roots from Eqs. (\ref{S67}) and (\ref{S68}) and hence get the GBZs in the complex plane. In the thermodynamic limit $M,N\rightarrow\infty$, the values of $\theta_\ell$ and $\phi_j$ will run continuously from $0$ to $\pi$, of course excluding several discrete complex values which correspond to the topological edge states, and thus the GBZs defined by Eq. (\ref{S66}) manifest as a circle in the complex plane, with a radius $|r|$ for $\beta_x$ and $|w|$ for $\beta_y$, as indicated in Supplementary Figs. \ref{figS6}(a) and \ref{figS6}(b).

\begin{figure}[h!]
\begin{center}
\includegraphics[width=16cm]{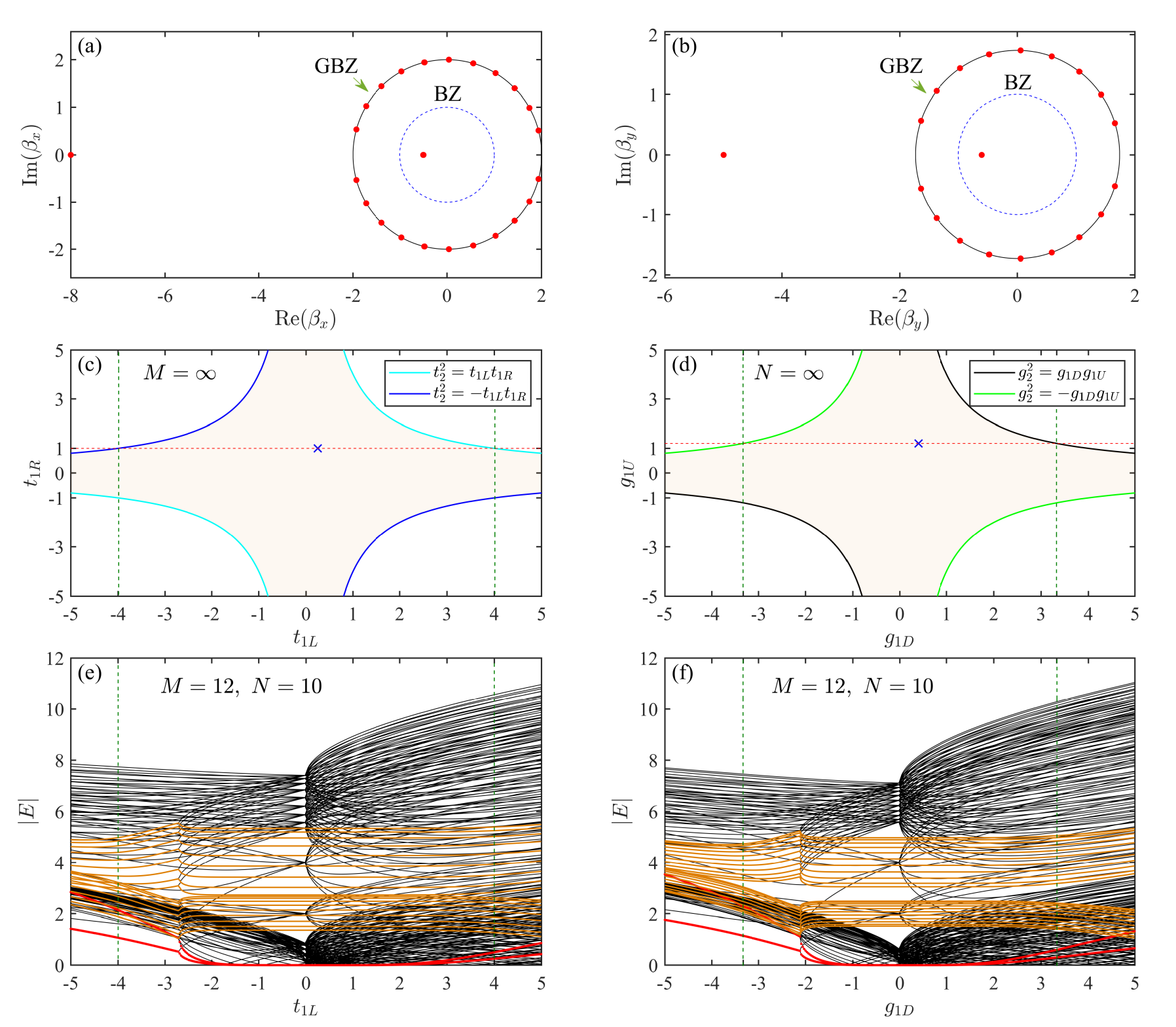}
\caption{(a,b) GBZs defined by $\beta_{x}=r\exp(\pm i\theta_\ell)$ and $\beta_{y}=w\exp(\pm i\phi_j)$, with red circles and solid curve corresponding to $M=12$ (or $N=10$) and the thermodynamic limit, respectively; (c,d) phase diagrams plotted in the ($t_{1L}, t_{1R}$) and ($g_{1D}, g_{1U}$) planes, wherein the blue cross corresponds to the parameter set used in Fig. 3 (right column) in the main text; and (e,f) the evolution of energy spectra $|E|$ for a finite lattice size of $M=12$ and $N=10$, along the red dashed line in phase diagrams, where black, yellow, and red lines denote the energy spectra of the bulk, edge, and corner states, respectively. All these plots use the same parameters as in Fig. 3 (right column) in the main text, unless noted otherwise.} \label{figS6}
\end{center}
\end{figure}

\subsection{Topological edge-state conditions and phase diagrams} \label{Sec3B}

As one can observe, the eigenenergy (\ref{S60}) depends on $R$ and $S$, but the latter two rely upon $\theta_\ell$ and $\phi_j$, respectively, which are in turn determined by the polynomial Eqs. (\ref{S67}) and (\ref{S68}).
Once the values of $\theta_\ell$ and $\phi_j$ are known for given set of hopping parameters, the eigenenergy (\ref{S60}) and the corresponding eigenstate given by Eqs. (\ref{S58}) and (\ref{S59}) are then obtained.

Particularly, for a 2D lattice of finite size $2M\times 2N$, when the hopping parameters satisfy the parameter condition
\begin{align}
|t_2/\sqrt{t_{1L}t_{1R}}|>1, \tag{S69}\label{S69}
\end{align}
the polynomial Eq. (\ref{S67}) will admit the asymptotic solution \cite{Hou23}
\begin{align}
\cos(\theta_\ell)\simeq\frac{\eta^{2M}(\eta^2-1)^2-\eta^2-1}{2\eta}, \tag{S70}\label{S70}
\end{align}
where $\eta=\sqrt{t_{1L}t_{1R}}/t_2$. In this situation, the $R$ formula (\ref{S47}) can now reduce to
\begin{align}
R\simeq \eta^{2M}(\eta^2-1)^2t_2^2. \tag{S71}\label{S71}
\end{align}
In an analogous manner, if the parameter condition
\begin{align}
|g_2/\sqrt{g_{1 D} g_{1 U}}|>1, \tag{S72}\label{S72}
\end{align}
is also fulfilled, Eq. (\ref{S68}) could possess the asymptotic solution
\begin{align}
\cos(\phi_j)\simeq\frac{\chi^{2N}(\chi^2-1)^2-\chi^2-1}{2\chi}, \tag{S73}\label{S73}
\end{align}
where $\chi=\sqrt{g_{1D}g_{1U}}/g_2$. Substituting Eq. (\ref{S73}) into Eq.~(\ref{S52}), one can obtain
\begin{align}
S\simeq \chi^{2N}(\chi^2-1)^2g_2^2. \tag{S74}\label{S74}
\end{align}
As seen from Eqs. (\ref{S71}) and (\ref{S74}), both $R$ and $S$ can approach zero in the thermodynamic limit $M,N\rightarrow\infty$ for some certain $\theta_\ell$ and $\phi_j$ values, provided that the parameter conditions (\ref{S69}) and (\ref{S72}) hold true. Therefore, the specific $\theta_\ell$ and $\phi_j$ values given by Eqs. (\ref{S70}) and (\ref{S73}) correspond to the topological edge states occurring in the horizontal and vertical chains, respectively.

Consequently, when only the parameter condition (\ref{S69}) is satisfied, the topological edge states pop up, with energies given by
\begin{align}
E^{\mathrm{dOBC}}_\mathrm{edge}=\pm\nu\sqrt{S},~~\mathrm{or}~~E^{\mathrm{dOBC}}_\mathrm{edge}=\pm\sqrt{S}, \tag{S75}\label{S75}
\end{align}
which are obtained, in the thermodynamic limit, from Eq. (\ref{S60}) by setting $R\rightarrow0$ therein, as indicated by the cyan and purple circles in Supplementary Fig. \ref{figS2}. Note that the corresponding $\theta_\ell$ value for these topological edge states is defined by Eq. (\ref{S70}). Likewise, if only the parameter condition (\ref{S72}) holds, the topological edge states would also emerge, with eigenenergies
\begin{align}
E^{\mathrm{dOBC}}_\mathrm{edge}=\pm\mu\sqrt{R}, ~~\mathrm{or}~~E^{\mathrm{dOBC}}_\mathrm{edge}=\pm\sqrt{R}, \tag{S76}\label{S76}
\end{align}
obtained from Eq. (\ref{S60}) by setting $S\rightarrow0$ therein. The corresponding $\phi_j$ value for such topological edge states is given by Eq. (\ref{S73}), while letting the $\theta_\ell$ values be determined by Eq. (\ref{S67}). Naturally, if both the parameter conditions (\ref{S69}) and (\ref{S72}) are met, the topological edge states exist and possess eigenenergies given by Eq. (\ref{S75}) or (\ref{S76}), as indicated by the cyan, purple, green, and pink circles in Supplementary Fig. \ref{figS3} or in Fig. 3 in the main text. Of course, in the last situation, when the $\theta_\ell$ and $\phi_j$ values are solely given by Eqs. (\ref{S70}) and (\ref{S73}), Eq. (\ref{S60}) becomes
\begin{align}
E^{\mathrm{dOBC}}_\mathrm{corner}=0, \tag{S77}\label{S77}
\end{align}
implying that the topological corner states appear, with a degenerate zero energy (see red circles in Supplementary Fig. \ref{figS3}).

In fact, the parameter conditions (\ref{S69}) and (\ref{S72}) for topological edge states can also be figured out from the GBZs based on Eqs. (\ref{S66}). Specifically, in the thermodynamical limit $M\rightarrow\infty$, Eq. (\ref{S70}) can be reduced to
\begin{align}
\theta_\ell^{\mathrm{edge}}\simeq \pi-\arccos\left[\frac{\eta^2+1}{2\eta}\right]. \tag{S78}\label{S78}
\end{align}
The value of $\theta_\ell^{\mathrm{edge}}$ is generally complex because of $\eta^2+1\geqslant 2\eta$. Only when $\theta_\ell^{\mathrm{edge}}$ is real does its $\beta_x^{\mathrm{edge}}$ locate precisely on the GBZ. Therefore, the borderline within which the topological edge states can appear in the $x$ direction is defined by $|\eta|=1$, viz.,
\begin{align}
|t_2/\sqrt{t_{1L}t_{1R}}|=1, \tag{S79}\label{S79}
\end{align}
completely consistent with the parameter condition (\ref{S69}). Similarly, from Eq. (\ref{S73}) and with the same arguments, one can find
\begin{align}
|g_2/\sqrt{g_{1D}g_{1U}}|=1, \tag{S80}\label{S80}
\end{align}
which defines the borderline for topological edge states occurring in the $y$ direction.

Based on Eqs. (\ref{S79}) and (\ref{S80}), one can plot the phase diagrams to show the regions where the topological edge states can exist in the thermodynamic limit. As an example, Supplementary Figs. \ref{figS6}(c) and \ref{figS6}(d) present the phase diagrams for topological edge states (see shaded area) in the ($t_{1L}, t_{1R}$) and ($g_{1D}, g_{1U}$) planes, respectively, with the other parameters being kept the same as in Figs. 3 (d,e,f) in the main text. Also, we demonstrate in Supplementary Figs. \ref{figS6}(e) and \ref{figS6}(f) the evolution of energy spectra $|E|$ along the red dashed lines in phase diagrams, where black, yellow, and red lines correspond to the energy spectra of the bulk states, the gapped edge states, and the topological corner states, respectively. It is seen that the phase diagrams can predict well the topological edge states for the $t_{1L}t_{1R}>0$ and $g_{1D}g_{1U}>0$ situations, but may give rise to large deviations for the $t_{1L}t_{1R}<0$ or $g_{1D}g_{1U}<0$ situation, due to the fact that a finite lattice size of $M=12$ and $N=10$ is adopted. The error for the latter case can be reduced significantly for sufficiently large $M$ and $N$ values used.